\documentclass[review]{elsarticle}

\usepackage{lineno,hyperref}
\usepackage{xspace}

\modulolinenumbers[5]

\usepackage{graphicx}
\usepackage{comment}
\usepackage{amsmath,amssymb} 
\usepackage{ulem}
\usepackage{enumerate}
\usepackage{CJK}
\usepackage{amsthm}
\usepackage{subfigure}
\usepackage{algorithm,algcompatible,amsmath}
\usepackage[noend]{algpseudocode}
\algnewcommand\INPUT{\item[\textbf{Input:}]}%
\algnewcommand\OUTPUT{\item[\textbf{Output:}]}%
\usepackage{amsmath}               
  {

  }

\newtheorem{definition}{Definition}

\pdfoutput=1

\begin{document}

\begin{frontmatter}

\title{Tree-based Search Graph for Approximate Nearest Neighbor Search}

\author{Xiaobin Fan \fnref{myfootnote1}}
\author{Xiaoping Wang\corref{cor1}\fnref{myfootnote1}}
\cortext[cor1]{Corresponding author}
\author{Kai Lu\fnref{myfootnote2}}
\author{Lei Xue\fnref{myfootnote1}}
\author{Jinjing Zhao\fnref{myfootnote3}}
\fntext[myfootnote1]{Hunan University, Changsha, 410082, China; E-mail: fanxbin@hnu.edu.cn, xpwang@hnu.edu.cn, xuelei@hnu.edu.cn}
\fntext[myfootnote2]{National University of Defense Technology, Changsha, 410073, China; E-mail: kailu@nudt.edu.cn}
\fntext[myfootnote3]{National Key Laboratory of Science and Technology on Information System Security, Beijing, China}

\begin{abstract}
 Nearest neighbor search supports important applications in many domains, such as database, machine learning, computer vision. Since the computational cost for accurate search is too high, the community turned to the research of approximate nearest neighbor search (ANNS). Among them, graph-based algorithm is one of the most important branches. Research by Fu et al. shows that the algorithms based on Monotonic Search Network (MSNET), such as NSG and NSSG, have achieved the state-of-the-art search performance in efficiency. The MSNET is dedicated to achieving monotonic search with minimal out-degree of nodes to pursue high efficiency. However, the current MSNET designs did not optimize the probability of the monotonic search, and the lower bound of the probability is only 50\%. If they fail in monotonic search stage, they have to suffer tremendous backtracking cost to achieve the required accuracy. This will cause performance problems in search efficiency. To address this problem, we propose (r,p)-MSNET, which achieves guaranteed probability on monotonic search. Due to the high building complexity of a strict (r,p)-MSNET, we propose TBSG, which is an approximation with low complexity. Experiments conducted on four million-scaled datasets show that TBSG yields better or comparable search performance than all recent state-of-the-art algorithms. Our code has been released on Github.
\end{abstract}

\begin{keyword}
information retrieval\sep approximate nearest neighbors search\sep graph-based search
\end{keyword}

\end{frontmatter}

\nolinenumbers

\section{Introduction}
Nearest neighbor search is to find the closest vector to a query vector from a given vector set. The distance measurement is defined by the application requirements, where most of them adopt euclidean distance. Nearest neighbor search is the fundamental technology in many domains including database, machine learning, and computer vision. However, exact nearest neighbor search in high dimensional space is often computationally expensive. Hence, researchers propose approximate nearest neighbor search (ANNS) to trade searching accuracy for computational efficiency\cite{arora2018hd}\cite{beis1997shape}\cite{zheng2016lazylsh}.

We classify the ANNS algorithm into four categories: tree-based methods\cite{bentley1975multidimensional}\cite{fukunaga1975branch}\cite{silpa2008optimised}, hashing-based methods\cite{gionis1999similarity}
\cite{huang2015query}\cite{weiss2008spectral}, quantization-based methods\cite{ge2013optimized}\cite{jegou2010product}\cite{zhang2014composite}, and graph-based methods\cite{harwood2016fanng}\cite{fu2019NSG}\cite{hajebi2011fast}\cite{DPG}\cite{malkov2014NSW}\cite{malkov2018HNSW}. Hashing-based methods try to map close vectors to the same bucket through multiple hash functions and reduce computation cost by only performing search in the bucket corresponding to the query vector. The hashing-based algorithms are renowned for Locality Sensitive Hashing (LSH)\cite{gionis1999similarity} and Spectral Hashing\cite{weiss2008spectral}. The tree-based methods recursively divide the vector set into multiple subsets until it meets the end condition. The establishment of tree index is often very fast, but too much backtracking is often required during search procedure, which leads to high computational cost. Typical methods include Randomized KD-Tree\cite{silpa2008optimised} and Ball Tree\cite{cayton2008fast}. The quantization-based methods try to decompose the original vector space into cartesian product of multiple subspaces and use the cartesian product of short codes instead of the original vector to perform the accelerated distance evaluation. Representative methods include Product Quantization (PQ)\cite{jegou2010product} and Composite Quantization (CQ)\cite{zhang2014composite}. Graph-based methods build a proximity graph that concatenates all vectors with vertexes representing vectors and edges representing proximity relationships. In query stage, the search starts from a fixed or randomly chosen node and moves to a neighbor closest to the query vector until it reaches the local optimum. A large number of experiments\cite{DPG}\cite{fu2019NSG}\cite{malkov2018HNSW}\cite{harwood2016fanng} show that the search efficiency of graph-based index outperforms the other three categories.

In recent years, graph-based ANNS attracts much research effort \cite{malkov2018HNSW}\cite{DPG}\cite{fu2019NSG}\cite{fu2021nssg}. Among them, algorithms based on Monotonic Search Network (MSNET) are in the leading position, such as NSG\cite{fu2019NSG} and NSSG\cite{fu2021nssg} . The current MSNET ensures that from any node \textit{A} to another node \textit{B} there is at least one path on which the distances from the nodes to node \textit{B} are monotonically decreasing. Since the query time cost can be roughly estimated as the product of length of search path and average out-degree of the nodes. The MSNET is dedicated to achieving monotonic search with minimal out-degree of nodes to pursue high efficiency. The monotonic search is that the search from any node \textit{A} always can find at least one neighbor of \textit{A} closer to query unless node \textit{A} is the nearest neighbor of the query. As the monotonic search prunes the unnecessary backtracking, the search path is much shorter. However, the current MSNET designs did not optimize the probability of the monotonic search, and the lower bound of the probability is only 50\%. If they fail in monotonic search stage, they have to suffer tremendous backtracking cost to achieve the required accuracy. This will cause performance problems in search efficiency.

\begin{figure}[!t]
\centering
\subfigure{\includegraphics[width=1.6in]{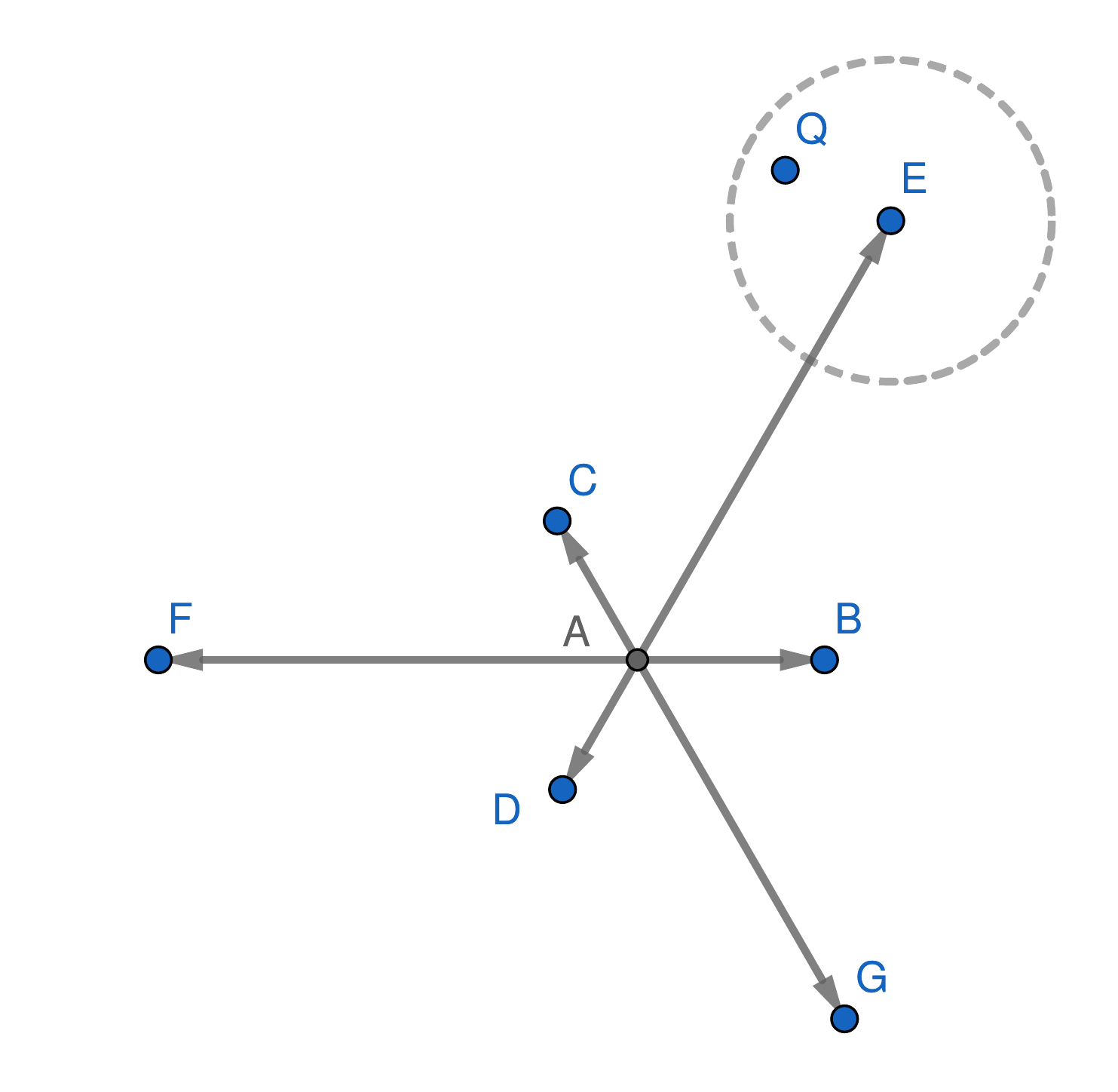}}
\subfigure{\includegraphics[width=1.6in]{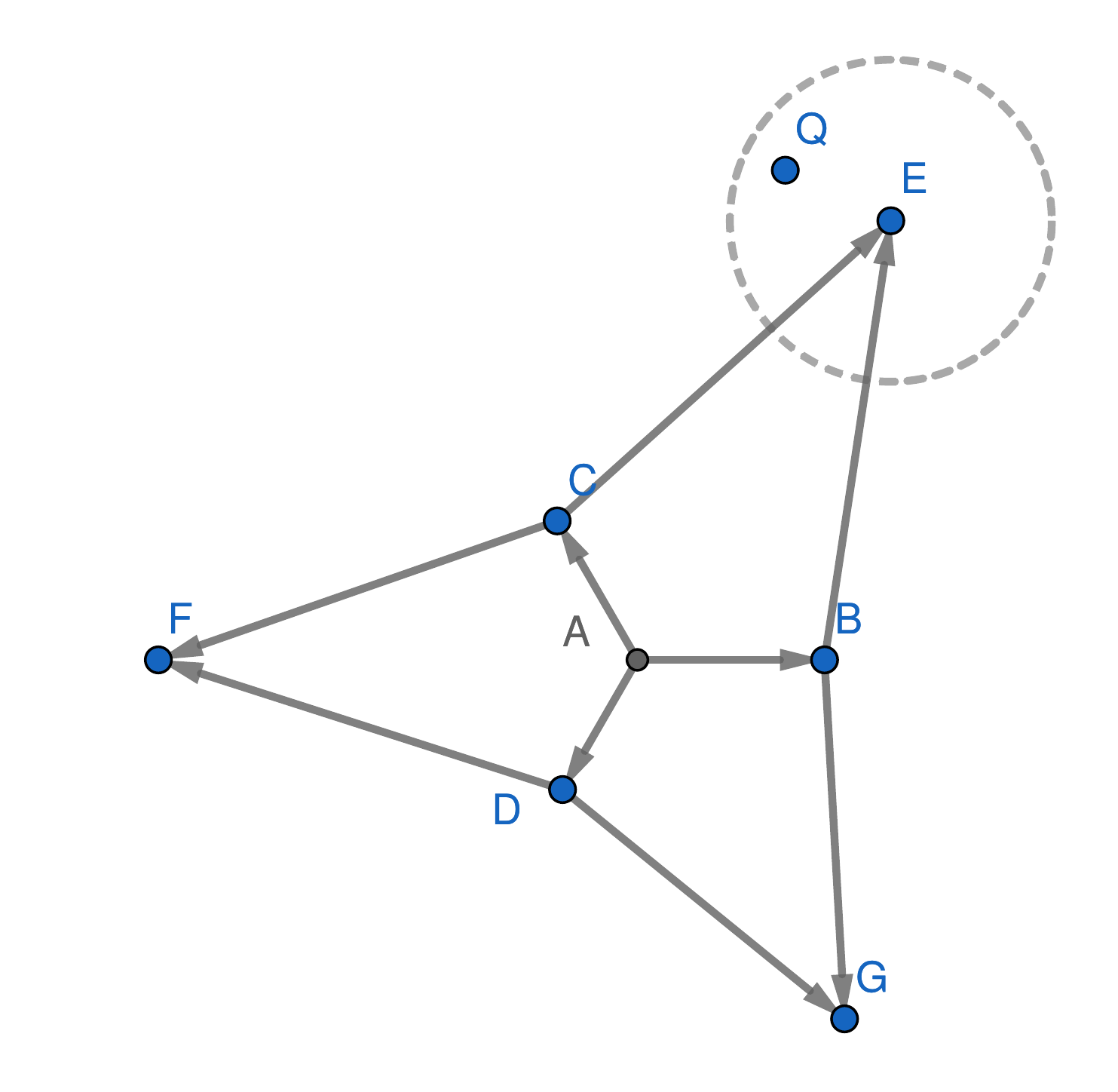}}
\caption{The left graph adopts NSSG 's strategy and point \textit{A} will connect all other points. While in the right graph point \textit{A} only connects points \textit{B}, \textit{C}, \textit{D}. For the search with \textit{Q} as query and starting from point \textit{A}, in left graph it needs to calculate the distance between 6 neighbors and \textit{Q}, while in right graph it will visit points \textit{B}, \textit{C}, \textit{D} and find point \textit{C} is closer to \textit{Q} and \textit{E}, \textit{F} are visited later. So in right graph it only need to evaluate 5 points-\textit{B}, \textit{C}, \textit{D}, \textit{E}, \textit{F}. }
\label{fig:disadvantage_of_NSSG}
\end{figure}

In this paper, we extent MSNET with probability guarantee, called $(r,p)$-MSNET. The probability of $(r,p)$-MSNET achieving monotonic search is at least \textit{p} when the distance between the nearest neighbor and the query point is less than \textit{r}. By adjusting the value of $p$, we can balance the length of the search path with the out-degree of nodes to further improve the search efficiency. The key to building a $(r,p)$-MSNET is the edge selection strategy. We prove that the lower bound of achieving monotonic search with the existing edge selection strategy is only 50\%. To address this problem, we propose a new edge selection strategy with greater lower bound.

However, the time complexity of building a strict $(r,p)$-MSNET is at least $O(n^2m)$ where \textit{n} is the size of nodes and \textit{m} is the size of neighbors, which is unacceptable. Therefore, in order to approximate a $(r,p)$-MSNET and reduce the complexity, we propose TBSG (Tree-based Search Graph). The index complexity of TBSG is close to $O(n)$ and it is applicable for many datasets. The detail of TBSG can be seen in Section 3.

The main contribution can be summarized as follows:
\begin{enumerate}[1.]
\item We propose the $(r,p)$-MSNET to match the query scenarios better.
\item We propose a new graph-based algorithm, called TBSG, which is an approximation of $(r,p)$-MSNET with low index complexity.
\item We conduct experiment on four million-scaled datasets. The result shows that TBSG yields better or comparable search performance than all recent state-of-the-art algorithms.
\end{enumerate}
\begin{CJK*}{UTF8}{gkai}
    \begin{algorithm}
        \caption{ SearchKnn($G, q, ep, l, k$)}
        \label{alg1}
        \begin{algorithmic}[1] 
        \Require Graph $G$,query point $q$, enter point $ep$, result pool size $l$, nearest neighbor size $k$
        \Ensure $k$ nearest neighbors to $q$
               \State result pool $ C \gets \phi $ 
               \State $ C.add(ep)$
                \While{$true$}
                \If{ nodes in $C$ are all visited}
                \State break
                \EndIf
                \State $c \gets $the first unvisited node in $C$  
                \State mark $c$ as visited
                \ForAll{neighbor $e$ of $c$ in $G$}
                \State $C.add(e)$
                \EndFor
                \State sort $C$ in the ascending order of the distance to $q$
                \If{$C.size()>l$}
                \State $C.resize(l)$
                \EndIf
                \EndWhile
                \State return the first $k$ nodes in C
        \end{algorithmic}
    \end{algorithm}
\end{CJK*}

\begin{table}
\centering
\caption{We list some frequently used notations here.}
\begin{tabular}{|c|c|} \hline
Notation&Definition\\ \hline
$\mathcal{D}$ & Dataset (point set) in $R^d$ \\ \hline
$n=\left |\mathcal{D}\right|$&Dataset cardinality\\ \hline
$d$& Dimension\\ \hline
$\delta(p_1,p_2)$& l2 distance between point $p_1$ and point $p_2$ \\ \hline
$\Pi_{\overrightarrow{v_1}}{\overrightarrow{v_2}}$& Projection of the vetor $v_2$ on vector $v_1$\\ \hline
\end{tabular}
\label{table:notation}
\end{table}
First we introduce the definition of ANNS, and then introduce the related work. For convenience, some frequently used notations are listed in Table \ref{table:notation}.
\subsection{Problem Setting}
We use $\mathcal{D}$ to denote a set of vectors in \textit{d} dimensional space $R^d$. Given a query point \textit{q}, a nearest neighbor query returns a point $o^* \in \mathcal{D}$ such that its Euclidean distance to q is the minimum among $\mathcal{D}$. Compared with exact NNS, the ANNS trades a little accuracy for efficiency. The ANNS problem can be defined as follows:
\begin{definition}
($\epsilon$-Nearest Neighbor Search).  Given a query point $q$ and an approximation ratio $\epsilon$,a  $\epsilon$-NNS query returns a point $o\in \mathcal{D}$ such that $\delta(o,q)\leq(1+\epsilon)\delta(o^*,q)$ where $o^* \in \mathcal{D}$ is the exact nearest neighbor of $q$.
\end{definition}
It is easy to generalize $\epsilon$-NNS to $\epsilon$-KNNS when we require the query return a sequence of $K$ points $<o_1,o_2,\ldots,o_k >$ such that for each $o_i(i \in [1,K])$, we have $\delta(o_i,q)\leq (1+\epsilon)\delta({o_i}^*,q)$ where ${o_i}^*$ is i-th nearest neighbor of $q$.
When we adopt proximity graphs to solve ANNS, it is difficult for us to model and evaluate results according to $\epsilon$. Instead, precision is used to evaluate the accuracy of searched results. Precision is defined as follows:
\begin{displaymath}precision=\frac{\left |R\cap G\right |}{\left |G\right |}\end{displaymath}where $R$ is result returned by algorithm and $G$ is groundtruth.
A higher precision indicates a smaller value of $\epsilon$, which means that the result is more accurate.

\subsection{Non-Graph-based Algorithm}
Research on non-graph-based algorithms is much earlier than that of graph-based algorithms. Many excellent algorithms are born among them, which can be roughly divided into three categories: hashing-based, tree-based and quantization-based methods. The hashing-based methods include LSH\cite{gionis1999similarity}, SH\cite{weiss2008spectral}, KLSH\cite{kulis2009kernelized}, USPLH\cite{wang2010sequential}, AGH\cite{liu2011hashing}, DSH\cite{jin2013density}, etc. They propose new hash functions to improve the probability of collisions between similar vectors and avoid collisions between non-similar vectors. The tree-based indexes include Kd-tree\cite{friedman1975algorithm}, R-tree\cite{beckmann1990r}, Cover Tree\cite{2015Faster}\cite{beygelzimer2006cover},  Ball-tree\cite{cayton2008fast}, etc. There are two main ways to search for tree-based index: backtracking on a tree or building multiple trees at the same time and querying the leaf node where the query vector is located. The quantization-based indexes include PQ\cite{jegou2010product}, OPQ\cite{ge2013optimized}, AQ\cite{babenko2014additive}, CQ\cite{zhang2014composite}, the goal of which is to minimize the quantization distortion.

\subsection{Graph-based Algorithm}
Graph-based Indexes contain a graph with \textit{n} nodes representing \textit{n} vectors and edges representing proximity relationship. Almost all the graph-based indexes adopt an algorithm similar to Algorithm \ref{alg1} for ANNS. We introduce them according their design discipline.

Navigable Small World Network (NSWN)\cite{boguna2009navigability}.
When performing ANNS on NSWN, the average length of searching path grows polylogarithmically with size of nodes. That is to say, the search complexity on NSWN is $O((\log{n})^v)$ where \textit{n} is the size of nodes and \textit{v} is a constant, which makes it quite suitable for ANNS. The NSW is an approximation of NSWN. The NSW\cite{malkov2014NSW} proposed by Y. Malkov et al achieves superior performance in experiment, largely due to its design: First, the process of inserting nodes is very similar to that of searching nodes. Second, the interconnections of newly inserted nodes and adjacent nodes are added to approximate the Delaunay Graph\cite{aurenhammer1991voronoi}. The first ensures logarithmic scaling of the greedy search while the second greatly improves the probability of finding the nearest neighbor. Later, they proposed Hierarchical NSW (HNSW)\cite{malkov2018HNSW} to solve the problem of decline in search efficiency caused by excessively large degree. The HNSW effectively controls the degree of nodes by leaving long edges on upper layers and using RNG's strategy to filter neighbors.

Monotonic Search Network (MSNET)\cite{dearholt1988monotonic}.The characteristic of Monotonic Search Network is to ensure at least one monotonic path from any node to another, making nodes on the path closer and closer to the target. The FANNG\cite{harwood2016fanng} approximates the MSNET by randomly selecting $k \times n(k \ll n)$ point pairs to meet the requirement of monotonic path, where \textit{k} is a hyper parameter and \textit{n} is size of nodes. The NSG\cite{fu2019NSG} combines points searched by the approximate KNNG from the navigating node and approximate k nearest neighbors as candidates. Then the RNG's strategy is used to select the neighbors from candidates, trying to ensure the existence of monotonous paths from the navigating node to all other nodes. Compared with NSG, the edge selection strategy adopted by NSSG\cite{fu2021nssg} is that if the angles between the candidate edge and all existing edges are greater than $\alpha(\alpha \leq 60^{\circ})$, the candidate edge will be selected, which effectively reduces the sparsity. Meanwhile, the NSSG directly uses approximate k nearest neighbors as the candidate set reducing the index complexity.

KNNG based graphs. The KGraph\cite{hajebi2011fast} directly uses the approximate K Nearest Neighbor Graph (KNNG) as the index. The DPG\cite{DPG} selects a fixed number of edges based on the approximate k nearest neighbors which maximizes the average pairwise angle. Then the reverse edges are added to enhance the connectivity. The Efanna\cite{fu2016efanna} uses Kd-tree as navigation and continue search on approximate KNNG.
\section{Algorithm and Analysis}
\subsection{Motivation}
Almost all the graph-based algorithms adopt algorithm like Algorithm \ref{alg1} for ANNS query. The algorithm can be simply summarized as: starting from a specific point, take it as the expanding point, calculate the distances between its neighbors and the query point then select the closest one as the next expanding point. Repeat it until the local optimum is met where no neighbor is closer to query than the expanding point. To approach the global optimum, we use a priority queue to store visited points for backtracking. The query time cost can be roughly estimated as the product of length of search path and average out-degree of the graph. In order to improve accuracy and efficiency, we put forward the definition of monotonic search.

\begin{definition}
(Monotonic Search).  For a given query point $q$, a graph defined on $\mathcal{D}$ achieves monotonic search if for each node $v_i \in G$ there are at least one edge $\overrightarrow{v_iv_j} \in G$ makes $\delta(q,v_j)<\delta(q,v_i)$ unless $v_i$ is the nearest neighbor of q.
\end{definition}

 We call the graph achieving monotonic search as Monotonic Search Network (MSNET). This definition of MSNET differs from MSNET in NSG\cite{fu2019NSG} and NSSG\cite{fu2021nssg}, which we will elaborate later. This MSNET is quite fit for nearest neighbor search, because it ensure that the search results are exact, and the search process does not need backtracking to ensure high efficiency. However, such an ideal MSNET is almost impossible to construct, mainly because the location of the query point $q$ cannot be predicted when we build the graph. To solve this problem, we propose the definition of $(r,p)$-MSNET.
 \begin{definition}
 ($(r,p)$-MSNET). For a given query point $q$, if $\delta(q,o^*)$ $\leq r$ where $o^*$ is the nearest neighbor of $q$, a graph is a $(r,p)$-MSNET when the probability of  achieving monotonic search is greater than $p$.
  \end{definition}
 For a given point set $\mathcal{D}$, we can construct a $(r,p)$-MSNET following three steps:
 \begin{enumerate}[1.]
\item First, we list all node $s \in \mathcal{D}$ as the starting node of search.
\item For each node $s$, we list all node $e \in E=\mathcal{D}-\{s\}$ as the nearest neighbor of query point $q$ assuming $\delta(q,e)$$ < r$. Sort $E$ in ascending order of $\delta(s,e)$. Here, $E$ is the candidate neighbor set of $s$ and $V=\phi$ originally.

\item For each node $e \in E$ we list all existing neighbor $v \in V$ and calculate the probability of $\delta(q,v) < \delta(q,s)$. If all the probabilities are lower than $p$, $e$ is added to $V$, otherwise not. Finally we set the neighbor set of $s$ in $G$ as $V$.
\end{enumerate}

It is easy to prove that the graph constructed by the above steps is a $(r, p)$-MSNET. When given a query point $q$ and the distance between $q$ and its nearest neighbor $o^*$ less than $r$, we start the search from any point $s$. If the point $s$ is not $o^*$, the point $o^*$ must be included in $E=\mathcal{D}-\{s\}$.  We use $V$ to denote the neighbor set of $s$, if there is a point $v \in V$ closer to query $q$ than $s$, the monotonic search can be achieved. Therefore, only if all the probabilities are lower than $p$, the edge $\overrightarrow{se}$ is necessary.  

\subsection{Edge selection strategy}
The key to build $(r,p)$-MSNET is the $3^{rd}$ step, which is called as edge selection strategy. In NSG, they define another type of MSNET which ensures from any point $A$ to another point $B$, there are at least one path on which the distance from the points to point $B$ decrease monotonically. It is worth noting that points $A$, $B$ must be in the point set $\mathcal{D}$. Therefore, the MSNET defined in NSG achieves monotonic search only if the query point is in the $\mathcal{D}$. However, in high dimensional space, this condition is difficult to guarantee, because the probability that the query point is not in $\mathcal{D}$ is much higher. The only difference between $(r,p)$-MSNET and MSNET defined in \cite{fu2019NSG} is the edge selection strategy.

The edge selection strategy adopted by NSG is the RNG' s strategy, which is also adopted by HNSW and FANNG. For a point $s$, the RNG's strategy will select the neighbor set $V$ from the candidate set $E$, which can be summarized as the following two steps:
 \begin{enumerate}[1.]
\item Sort $e \in E$ in the ascending order of $\delta(s,e)$;
\item For each point $e \in E$, only if all the point $v \in V$ satisfies $\delta(s,e)<\delta(v,e)$, $e$ will be added to $V$,  otherwise not.
\end{enumerate}

\begin{figure}
\centering
\includegraphics[width=1.6in]{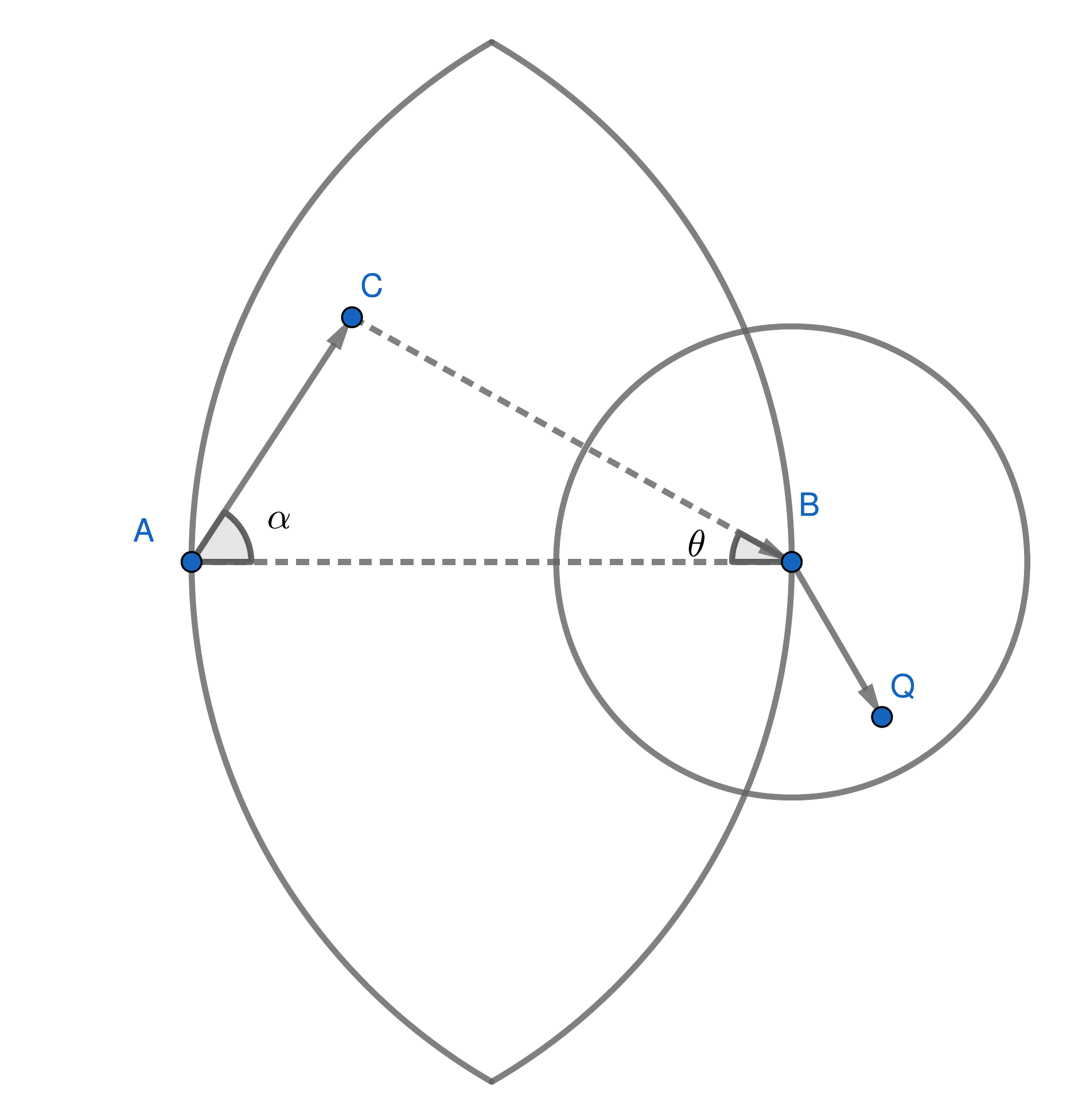}
\caption{point \textit{A} is the current node, where point \textit{C} is one of existing neighbors, point \textit{B} is the candidate neighbor and point \textit{Q} is the query point. Also, point \textit{B} is the nearest neigbhor of \textit{Q}.}
\label{fig:RNG_strategy}
\end{figure}

In other words, the point $s$ will not connect a candidate point if there is an existing neighbor of $s$ closer to the candidate point. We prove that when the query point is not in $\mathcal{D}$, the RNG's strategy makes the lower bound of the possibility of achieving monotonic search only 50\%. We demonstrate it in combination of Figure \ref{fig:RNG_strategy}. Point \textit{A} is the current node which is selecting neighbors. Point \textit{C} is point A 's one of the existing neighbors, point \textit{B} is the candidate neighbor and point \textit{Q} is the query point. At the same time, point \textit{B} is the nearest neighbor of \textit{Q}. According to RNG's strategy, point \textit{A} will not connect point \textit{B} because $\delta(C,B) < \delta(A,B)$. According to the definition of monotonic search, the condition to achieving monotonic search from point \textit{A} is $\left | \overrightarrow{CQ} \right| <\left| \overrightarrow{AQ}\right | $. If point \textit{Q} $\in \mathcal{D}$, then point \textit{B} equals point \textit{Q}. There must be $\left | \overrightarrow{CQ} \right| <\left| \overrightarrow{AQ}\right | $, the possibility is 100\%. However, if point \textit{Q} $\notin \mathcal{D}$ , we can see that 
\begin{displaymath}\left | \overrightarrow{AQ} \right| >\left| \overrightarrow{CQ}\right | \end{displaymath}
\begin{displaymath} \Pi_{\overrightarrow{AC}}\overrightarrow{AQ}> \frac{1}{2}\left | \overrightarrow{AC} \right |\end{displaymath}
\begin{displaymath}\Pi_{\overrightarrow{AC}}\overrightarrow{AB} + \Pi_{\overrightarrow{AC}}\overrightarrow{BQ} > \frac{1}{2}\left | \overrightarrow{AC} \right | \end{displaymath}
\begin{equation}\label{equa1} \Pi_{\overrightarrow{AC}}\overrightarrow{BQ} > \frac{1}{2}\left | \overrightarrow{AC} \right | - \Pi_{\overrightarrow{AC}}\overrightarrow{AB}\end{equation}

The $\Pi_{\overrightarrow{AC}}\overrightarrow{AQ}$ represents the projection of the vector $\overrightarrow{AQ}$ on the vector $\overrightarrow{AC}$. For convenience for discussion, we set that $l=\left | \overrightarrow{AB} \right |$, $\alpha=\angle CAB$, $\theta=\angle ABC$. Inequation (\ref{equa1}) can be represented as
\begin{displaymath}\Pi_{\overrightarrow{AC}}\overrightarrow{BQ} >\frac{l\sin\theta}{2 \sin(\alpha+\theta)} - l\cos\alpha\end{displaymath}
\begin{equation} \label{equa3}\Pi_{\overrightarrow{AC}}\overrightarrow{BQ} >-\frac{l\sin(2\alpha+\theta)}{2\sin(\alpha+\theta)}\end{equation}
Because $0< \left | \overrightarrow{AC} \right |,\left | \overrightarrow{CB} \right | \leq \left | \overrightarrow{AB} \right | $, there is $0\leq\alpha, \theta\leq\pi-\alpha-\theta$, and $0\leq\alpha+\theta\leq2\alpha+\theta\leq\pi$. It is easy to show that$-\frac{l\sin(2\alpha+\theta)}{2\sin(\alpha+\theta)}\leq 0$. That is to say, inequation (2) holds as long as the angle between $\overrightarrow{AC}$ and $\overrightarrow{BQ}$ $\leq 90^\circ$. According to the symmetry of the hyper sphere, the possibility is not less than 0.5 and equal to 0.5 if $2\alpha+\theta=\pi$. Therefore, the lower bound is 50\%. 

In NSSG\cite{fu2021nssg}, Fu et al. propose a new edge selection strategy. For a point $s$, the NSSG's strategy will select the neighbor set $V$ from the candidate set $E$ with threshold $\alpha_t$ $(\alpha_t \leq 60^{\circ})$ following two steps:
 \begin{enumerate}[1.]
\item Sort $e \in E$ in the ascending order of $\delta(s,e)$;
\item For each point $e \in E$, only if all the point $v \in V$ satisfies $\angle vse > \alpha_t$, $e$ will be added to $V$,  otherwise not.
\end{enumerate}

That is to say, point \textit{A} will not connect \textit{B} if $\alpha \leq \alpha_t$. If $\alpha_t=60^\circ$, according to inequation (2), the lower bound is also only 50\%. If $\alpha_t<60^\circ$, it will cause the over large out-degree of nodes, which lead to decline in search efficiency. We illustrate this by a toy example in Figure \ref{fig:disadvantage_of_NSSG}.

\subsection{ A new edge selection strategy}
In order to construct a $(r,p)$-MSNET, we expect the edge selection strategy to guarantee the  probability for any given value. Therefore, the key of the edge selection strategy is the calculation of the probability. 
\begin{figure}
\centering
\includegraphics[width=2.5in]{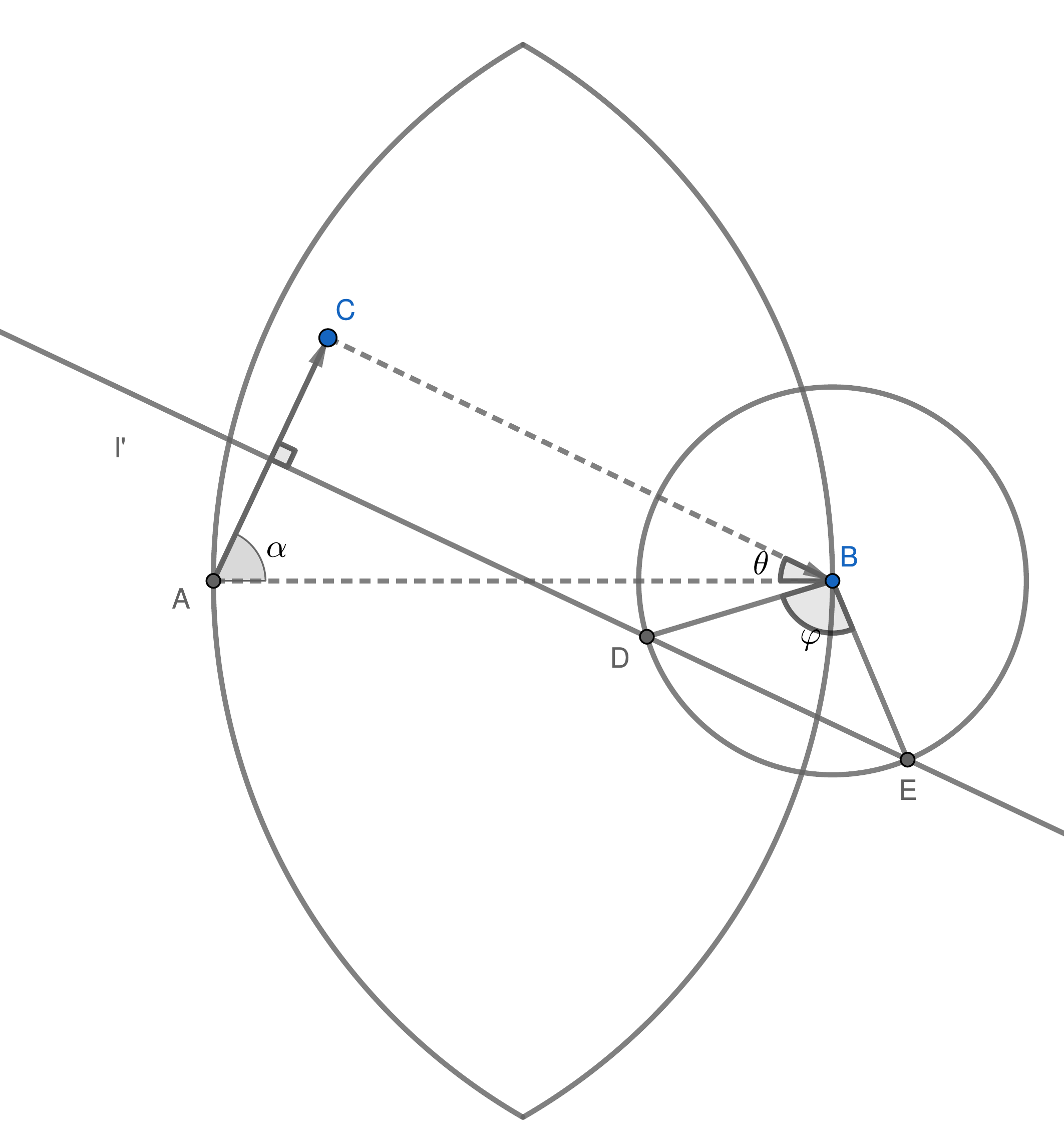}
\caption{point \textit{A} is the current node, where point \textit{C} is one of existing neighbors, point \textit{B} is the candidate of point \textit{A}. Point \textit{B} is the nearest neighbor of the point \textit{Q}. }
\label{fig:Min_prob}
\end{figure}

We illustrate the possibility in combination with Figure \ref{fig:Min_prob}. First we consider the case in two dimensional space, and the relevant conclusions can be extended to higher dimensional space. Point \textit{A} is the current point selecting neighbors, where point \textit{C} is one of existing neighbors, point \textit{B} is the candidate neighbor. Line \textit{l'} is the line segment $\overline{AC}$ 's perpendicular bisector. Line \textit{l'} intersects the circle with node \textit{B} as center and \textit{r} as radius at points \textit{D} and \textit{E}. $\varphi = \angle DBE$. Line $l'_B$ is parallel to line $l'$ and passes through the point B. We can show that
\begin{displaymath}distance(l',B) = distance(l',l'_B) 
= distance(l'_B,A)-distance(l',A)\end{displaymath} \begin{displaymath}=\Pi_{\overrightarrow{AC}}\overrightarrow{AB}-\frac{1}{2}\left | \overrightarrow{AC} \right |  =\frac{l\sin(2\alpha+\theta)}{2\sin(\alpha+\theta)}\end{displaymath}

The line \textit{l'} divides the circle into two parts: upper part and lower part. When the query point \textit{Q} locates at the upper part, since $\delta(C,Q)<\delta(A,Q)$, the monotonic search from point \textit{A} can continue. However, when the query \textit{Q} locates at the lower part, the monotonic search will fail without the edge $\overrightarrow{AB}$. That is to say, the probability of achieving monotonous search from node \textit{A} can be represented as the proportion of the area (volume) of the upper part. In high dimensional space, the line \textit{l'} is a hyperplane. The circle with point \textit{B} as center and \textit{r} as radius becomes a hypersphere. If we set up a cartesian coordinate system $x_1,x_2,\ldots,x_d$ in \textit{d} dimensional space with point \textit{B} as origin and we specify that the positive direction of axis $x_1$ is same as $\overrightarrow{AC}$. The probability can be represented as
\begin{equation}\frac{ \int_{-r\cos{\frac{\varphi}{2}}}^{r} d x_1\int_{-\sqrt{r^2-{x_1}^2}}^{\sqrt{r^2-{x_1}^2}}d x_2 \ldots \int_{-\sqrt{r^2-{x_1}^2-{x_2}^2-\ldots-{x_{d-1}}^2}}^{\sqrt{r^2-{x_1}^2-{x_2}^2-\ldots-{x_{d-1}}^2}}d x_d}{ \int_{-r}^{r} d x_1\int_{-\sqrt{r^2-{x_1}^2}}^{\sqrt{r^2-{x_1}^2}}d x_2 \ldots \int_{-\sqrt{r^2-{x_1}^2-{x_2}^2-\ldots-{x_{d-1}}^2}}^{\sqrt{r^2-{x_1}^2-{x_2}^2-\ldots-{x_{d-1}}^2}}d x_d}\end{equation}

\begin{figure}
\centering
\includegraphics[width=2in]{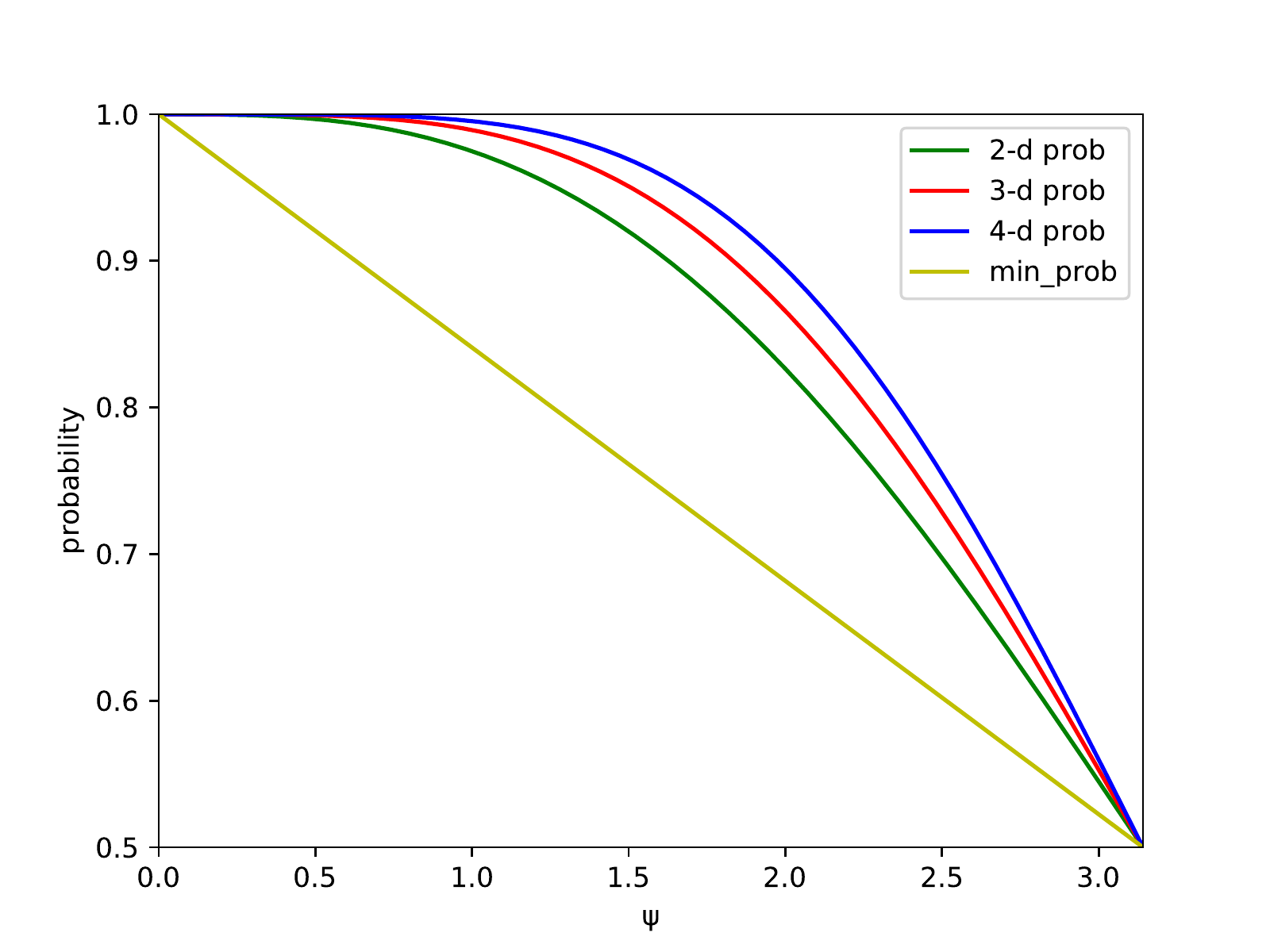}
\caption{Comparison between probabilities in 2, 3, 4 dimensional space and min\_prob }
\label{prob_comp}
\end{figure}

The probability is difficult to calculate in high dimension, but we prove that it must not be less than $1-\frac{\varphi}{2\pi}$. Therefore we define
\begin{equation}\label{equa2}min\_prob=1-\frac{\varphi}{2\pi}=1-\frac{\arccos{(\frac{l}{r}\frac{\sin{(2\alpha+\theta)}}{2\sin{(\alpha+\theta)}})}}{\pi}\end{equation}
We calculate the probabilities in 2, 3 and 4 dimensional spaces and we can see how they vary with $\varphi$ in Figure \ref{prob_comp}. With the increase of dimension, the gap between accurate probability and min\_prob tends to increase, so min\_prob is a loose lower bound. Also, both probabilities and the min\_prob decrease monotonically with the increase of $\varphi$, we can adjust the probability through min\_prob. 

Our edge selection strategy is based on min\_prob . For a node \textit{s}, with \textit{E} as candidate set and \textit{mp} as threshold of min\_prob, we perform the following steps to select neighbor set \textit{V}:
\begin{enumerate}[1)]
\item Sort $e \in E$ in the ascending order of $\delta(s,e)$;
\item For each node $e \in E$ we list all existing neighbor $v \in V$ with $\delta(v,e)<\delta(s,e)$ and calculate the min\_prob of monotonic search with edge $\overrightarrow{sv}$. If all are lower than $mp$, $e$ is added to $V$, otherwise not.
\end{enumerate}

\begin{figure}[!t]
\centering
\subfigure[RNG 's strategy]{\includegraphics[width=1.5in]{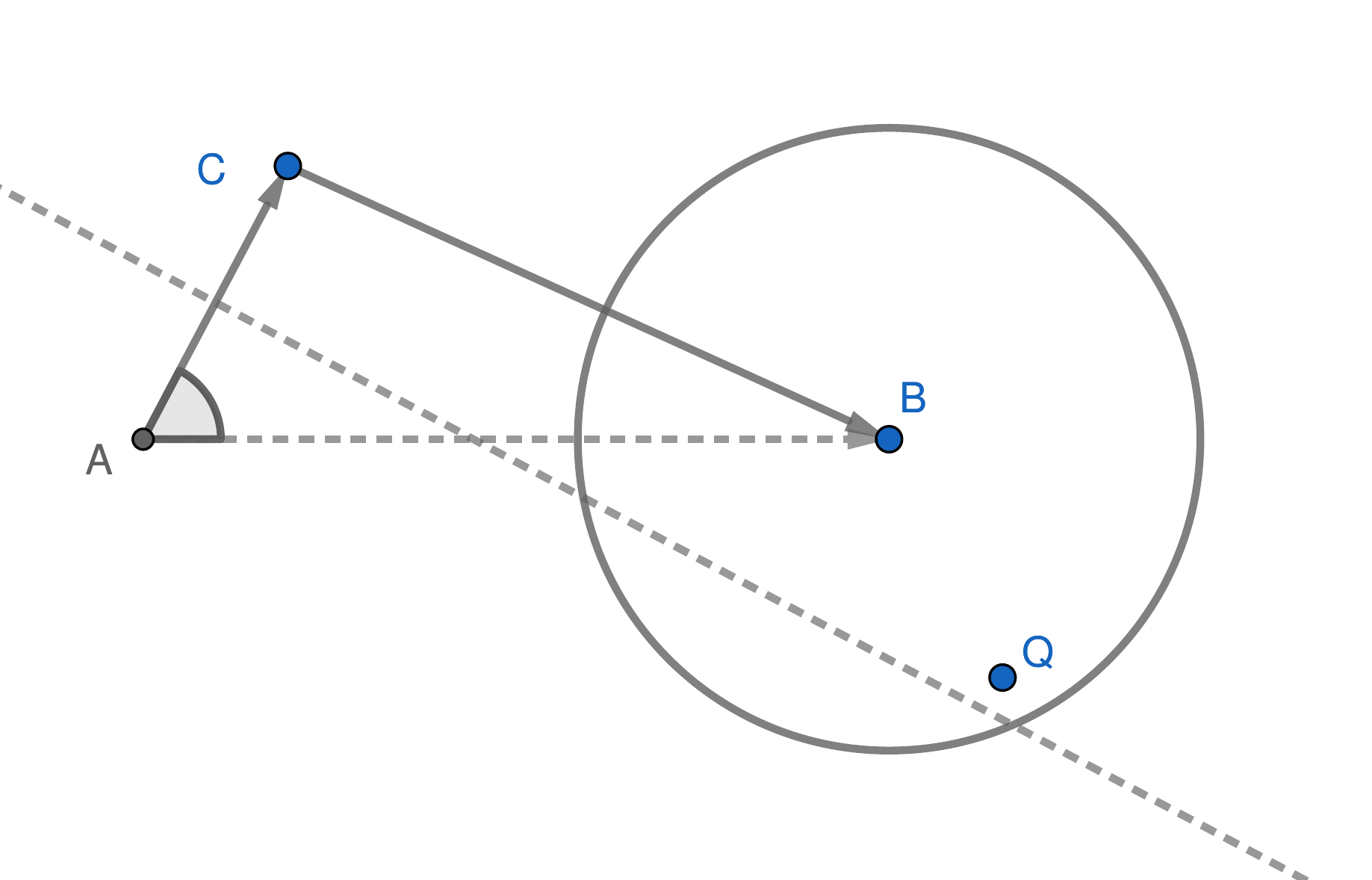}}
\subfigure[RNG 's strategy]{\includegraphics[width=1.5in]{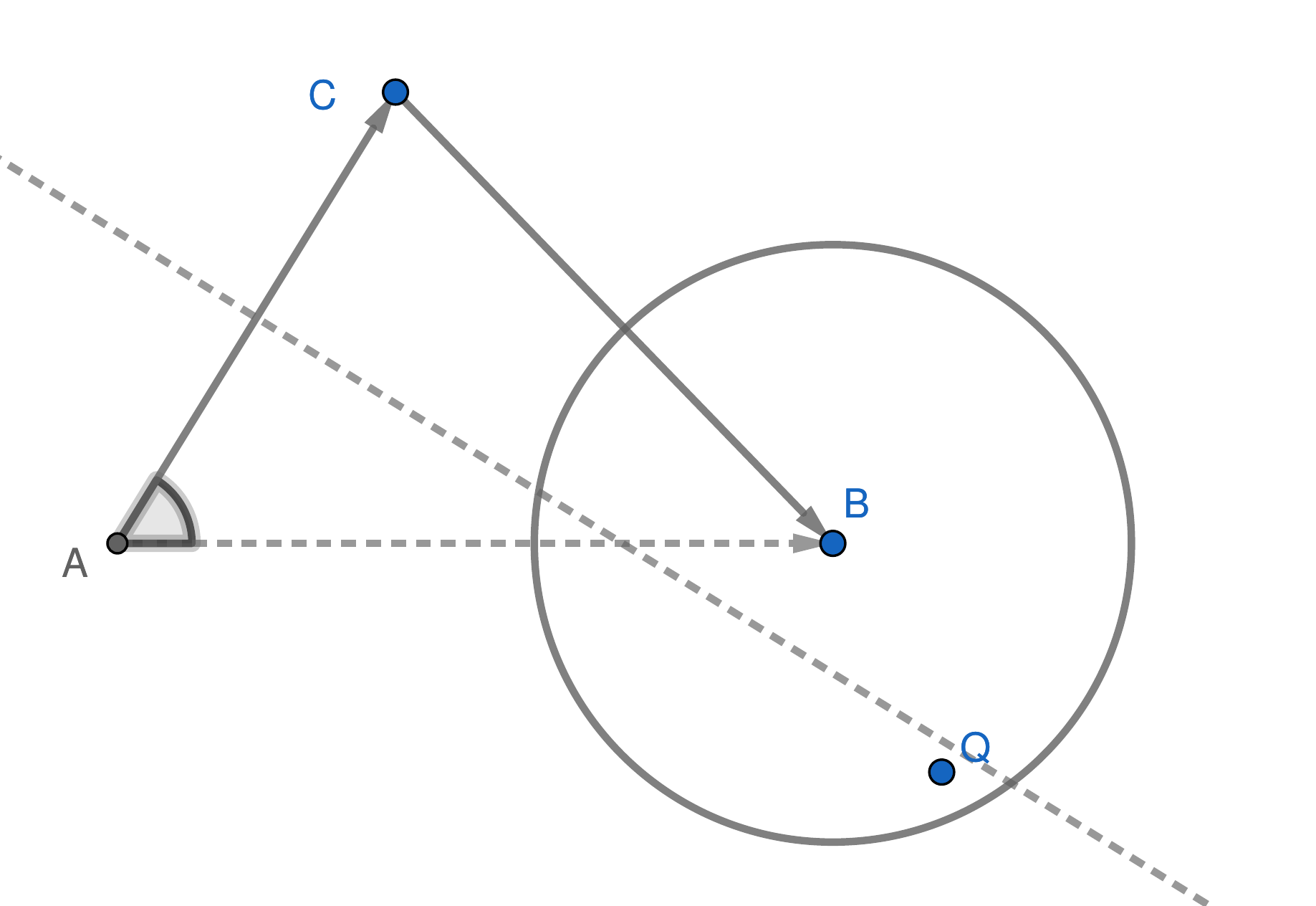}}
\subfigure[NSSG 's strategy]{\includegraphics[width=1.5in]{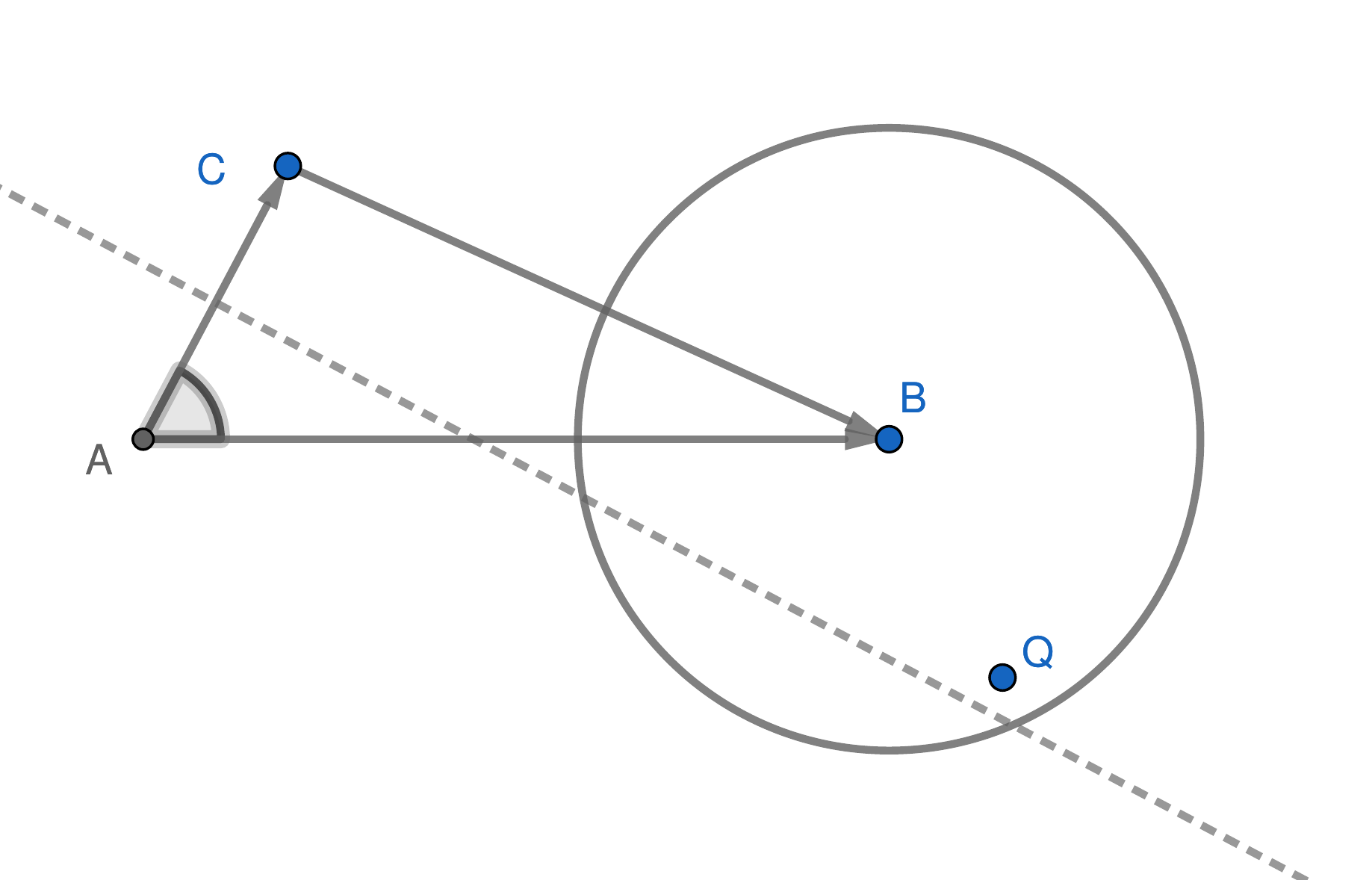}}
\subfigure[NSSG 's strategy]{\includegraphics[width=1.5in]{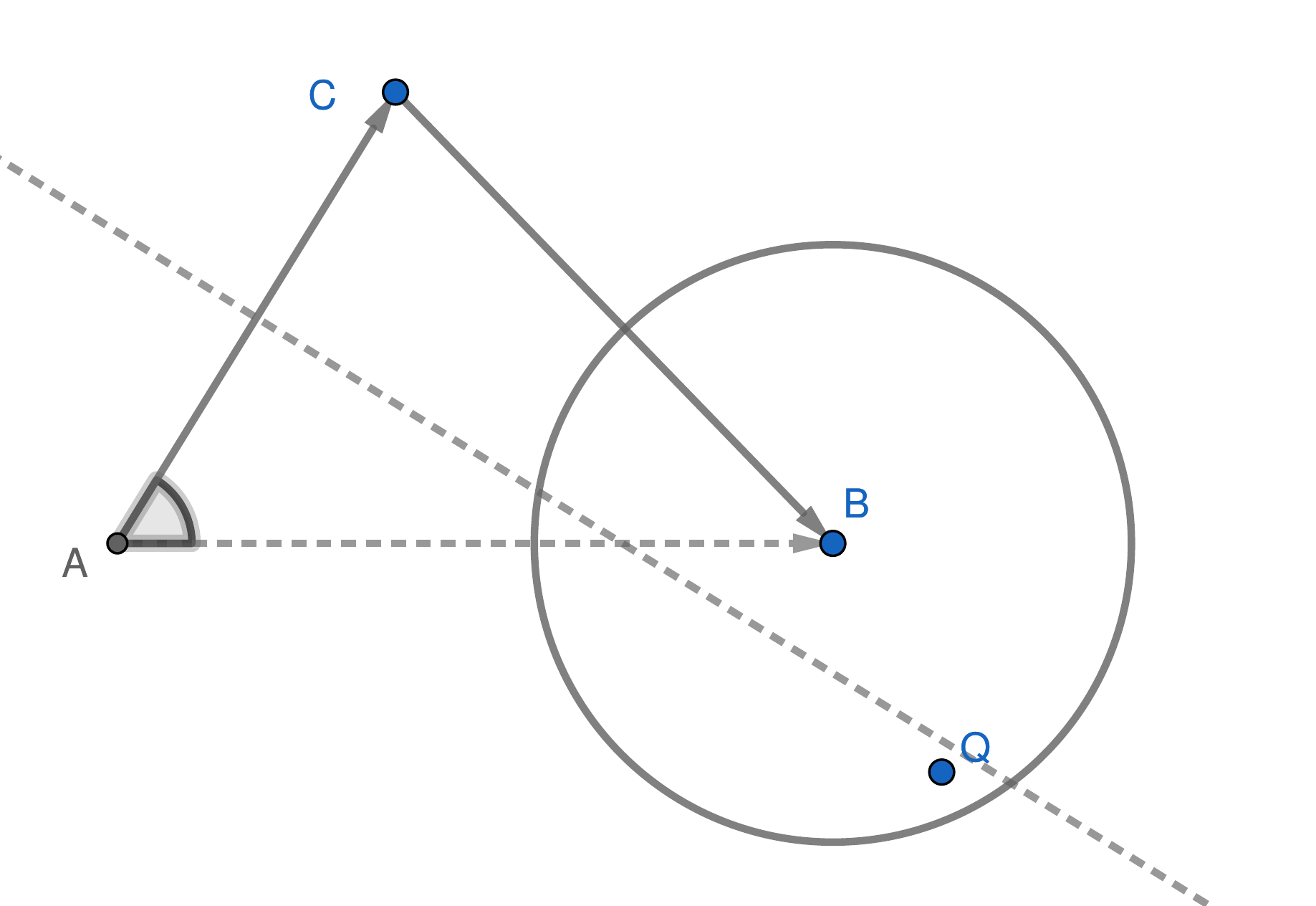}}
\subfigure[TBSG 's strategy]{\includegraphics[width=1.5in]{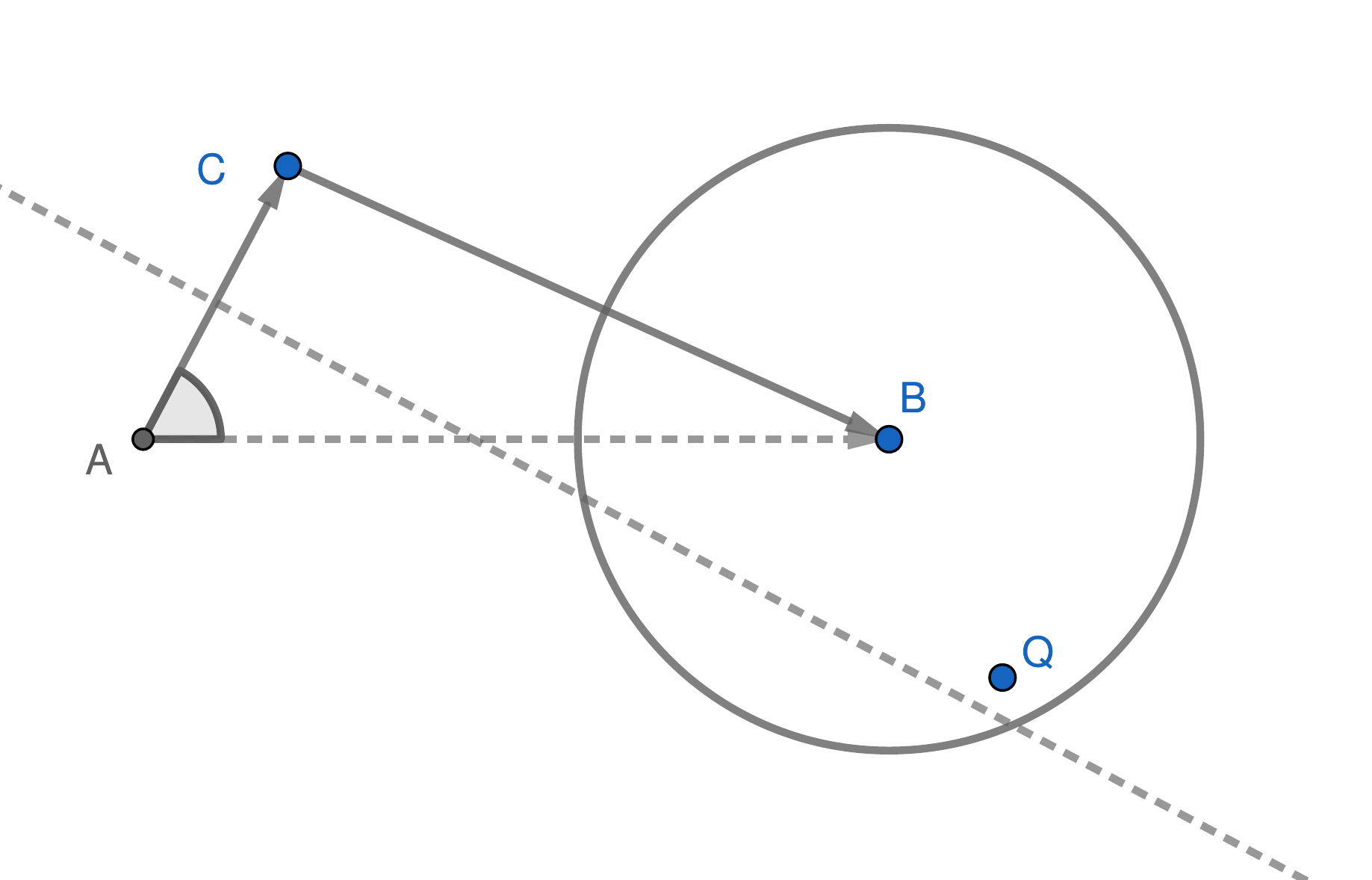}}
\subfigure[TBSG 's strategy]{\includegraphics[width=1.5in]{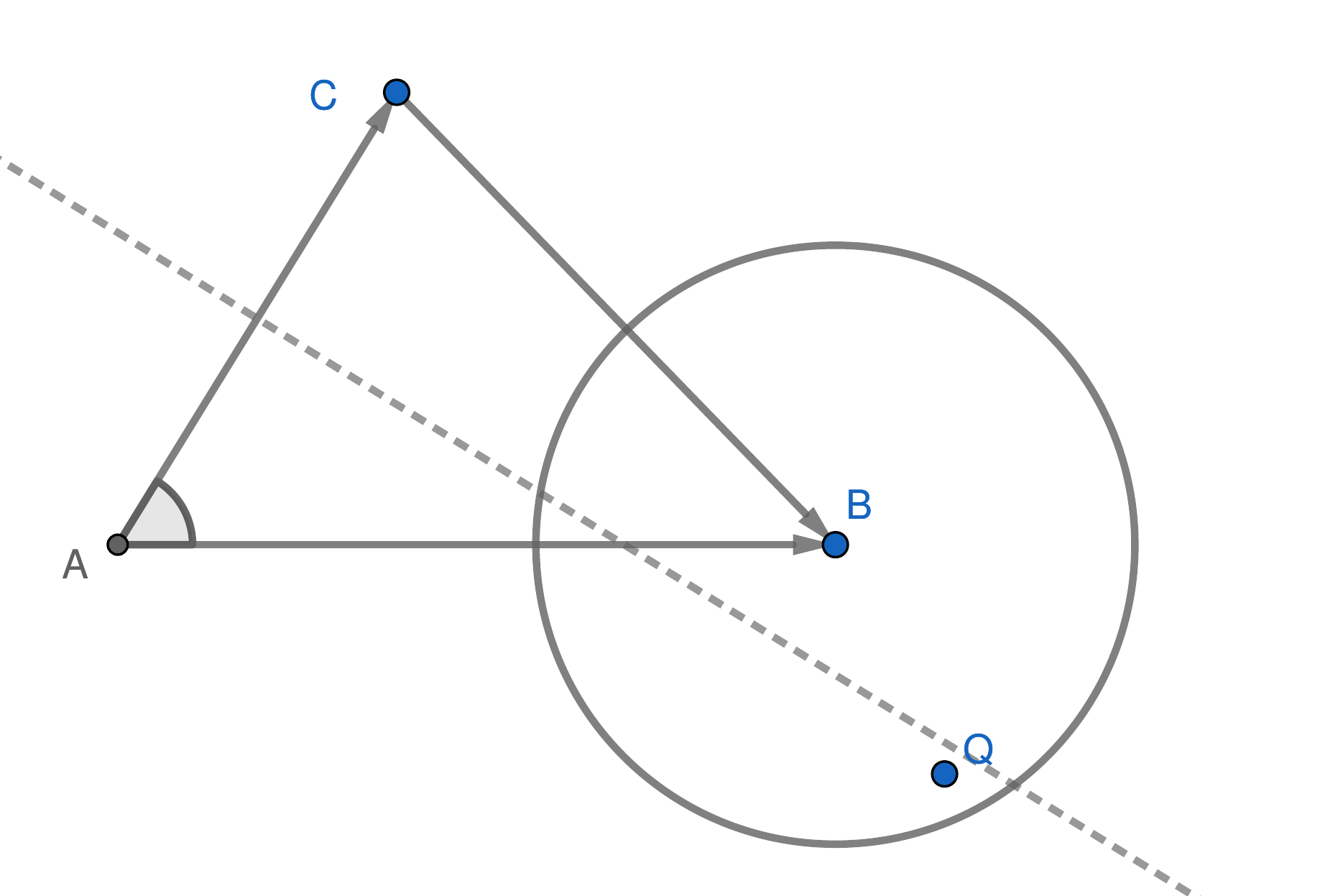}}
\caption{ The solid vectors represent connections and the dashed represent disconnections. In (a)(c)(e), with $\angle CAB=62.1^\circ$, and \textit{Q} as query, because node \textit{C} is closer to \textit{Q} than node \textit{A}, the search from node \textit{A} can reach node \textit{B} through node \textit{C}. However, in (b)(d)(f), with $\angle CAB=58.3^\circ$ , because node \textit{C} is further to \textit{Q} than node \textit{A}, connection from node \textit{A} to node \textit{B} is necessary. Because the min\_prob in (e) is greater than that in (f), we can remove $\protect \overrightarrow{AB}$ in (e) but reserve in (f) by setting a median threshold.  It shows that our strategy is more effective in selecting connections that are necessary.}
\label{strategy_compare}
\end{figure}

There is a simple example showing the difference between three popular edge selection strategies in Figure \ref{strategy_compare}. To further improve the search efficiency, we expect to achieve monotonic search with guranteed probability meanwhile keeping smallest out-degree.

\subsection{TBSG}
To build a $(r,p)$-MSNET, for each node its candidate neighbor set should include all other nodes. Therefore, the time complexity is at least $O(n^2m)$, where $n$ is the cardinality of  the dataset and $m$ is the maximum of neighbor size. An effective way to reduce the complexity is that for each node its candidate neighbor set only includes nodes close to it. It is effective especially for $(r,p)$-MSNET. According to the equation (\ref{equa2}), when $l$ is much greater than $r$, the min\_prob is going to be large. Therefore, it does not make much sense to add nodes far away to the candidate set.

However, for the datasets that consist of multiple separate clusters, taking  the nearest neigbhors set as candidate may result in the loss of global connectivity. Therefore, in order to construct a $(r,p)$-MSNET with global connectivity, we propose TBSG (Tree-based Search Graph), which consists of two parts: Cover Tree  and BKNNG (Bi-directed K Nearest Neighbor Graph). The Cover Tree is a data structure for exact NNS, which recursively divides a spherical space into several smaller spherical spaces and owns strong connectivity.
The BKNNG is a graph in which each node has bidirectional connections with its \textit{K} nearest neighbors. The value of \textit{K} ranges from hundreds to thousands. The TBSG combines the both and adopts our edge selection strategy to select neighbors to approximate a $(r,p)$-MSNET.

\subsubsection{Cover Tree}
Cover Tree, proposed by Beygelzimer et al.\cite{beygelzimer2006cover}, is a data structure used for exact nearest neighbor search. It recursively divides a spherical space into several smaller spherical spaces, each with a vector as the center, and the radius of the sphere decreases exponentially with the depth of the tree. Izbicki et al.\cite{2015Faster} simplified the Cover Tree structure to make the number of nodes in the tree exactly equal to size of dataset, which is also the Cover Tree used in this paper.
\begin{figure}
\centering
\includegraphics[width=2.5in]{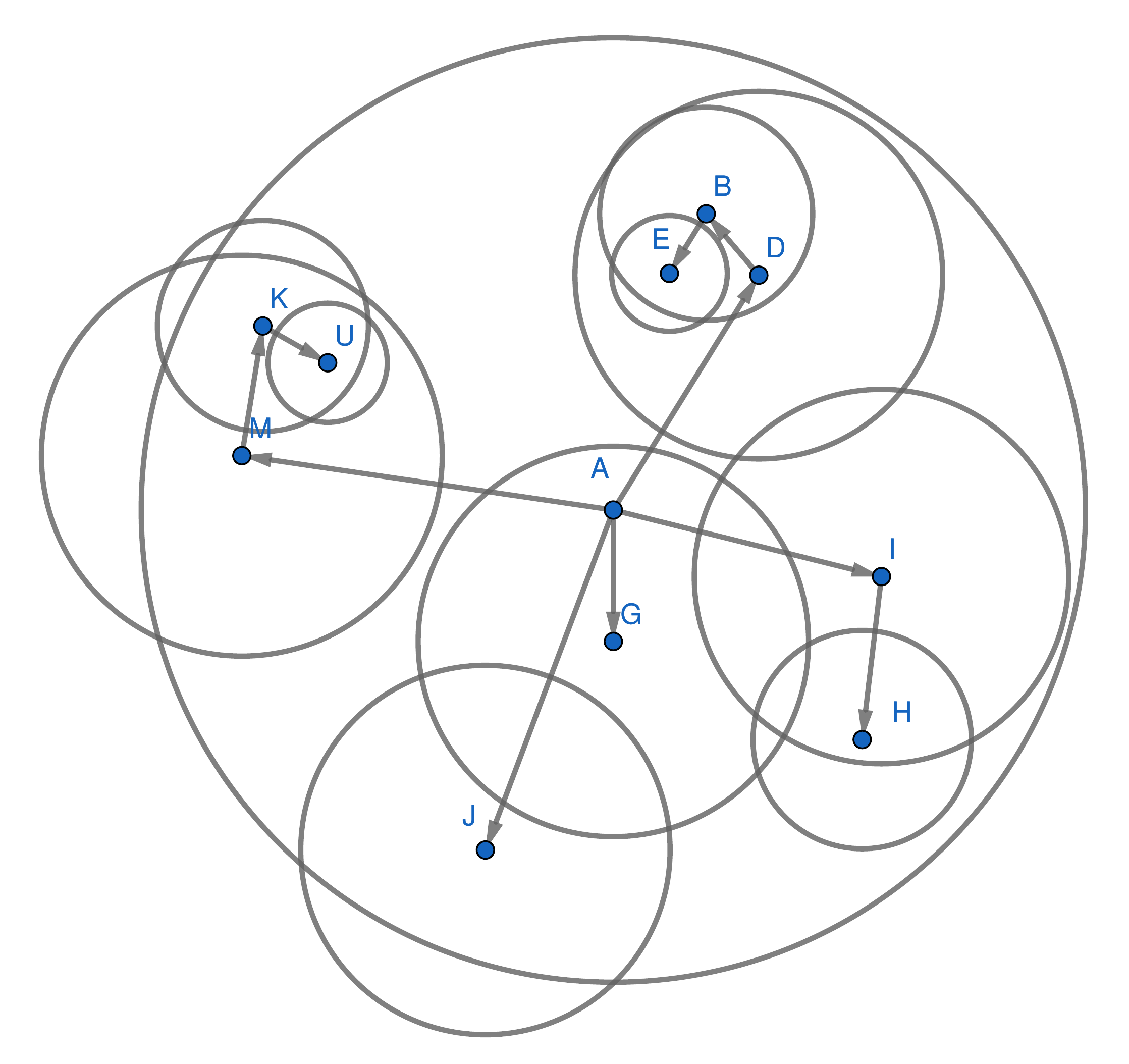}
\caption{A toy example of Cover Tree in  two dimensional space.}
\label{fig:Cover_tree}
\end{figure}
\newline
We present a two-dimensional Cover Tree example in Figure \ref{fig:Cover_tree}. To construct a Cover Tree, first we select a node $e_{root}$ as the root node, and then we insert other nodes into the Cover Tree. To insert a new node, we start from $e_{root}$ and search for the node $e_{child}$ that is closest to the inserted node among the children of  $e_{root}$. If the inserted node is in the sphere space of the node $e_{child}$, then insert into the subtree of $e_{child}$ recursively. Otherwise, add a new subtree of $e_{root}$ with the inserted node as root.

\begin{CJK*}{UTF8}{gkai}
    \begin{algorithm}
        \caption{ NeighborSelection($s,E,m,mp,r$)}
        \label{algo3}
        \begin{algorithmic}[1] 
        \Require node $s$ selecting neighbors, neighbor candidate set $E$, maximum of neighbor size $m$, threshold of min\_prob $mp$, radius r
        \Ensure selected neighbor set V
             \State $V \gets \phi$
             \State sort $E$ in the ascending order of distance to s
             \ForAll {node $e$ in $E$}
             \If{$V.size()==m$} break
             \EndIf
             \State exclude $\gets$ false
             \ForAll {node $v$ in $V$}
             \If{$\delta(v,e)<\delta(s,e)$}
              \State $m\_prob \gets $ calculate the min\_prob with edge $\overrightarrow{sv}$
              \If{$m\_prob \ge mp$} 
                 \State exclude $\gets$ true
                 \State break
              \EndIf
             \EndIf
             \EndFor
              \If{not exclude} $V.add(e)$
              \EndIf
              \EndFor
              \State return $V$
        \end{algorithmic}
    \end{algorithm}
    \end{CJK*}

\begin{CJK*}{UTF8}{gkai}
\begin{algorithm}
        \caption{ TBSG\_Construction($CT, KG, m, mp, r, n$)}
        \label{algo4}
        \begin{algorithmic}[1] 
        \Require Cover Tree $CT$, KNNG $KG$, maximum of neighbor size $m$, threshold of min\_prob $mp$, radius $r$, dataset cardinality $n$
        \Ensure TBSG $G$ with enter point $ep$
       \State $ep \gets$ root of CT
       \State $G \gets $Graph with $n$ nodes and no edges
       \State BG $\gets$ add reversed edges to KG
        \ForAll{node $s$ in $G$}
       \State $E \gets$ neighbors of $s$ in $BG$
       \State $E .add($s.children()$)$ in $CT$    
        \State $V \gets $NeighborSelection($s,E,m,mp,r$)
        \State set the neighbors of $s$ in $G$ as $V$
             \EndFor
             \State return G
        \end{algorithmic}
    \end{algorithm}
    \end{CJK*}
\subsubsection{TBSG}
In order to approximate a $(r, p)$-MSNET and reduce the time complexity, we need to construct a KNNG (K Nearest Neighbor Graph) of high quality and efficiency. There are many algorithms for fast construction of approximate KNNG, we adopt the algorithm in Efanna\cite{fu2016efanna}, the complexity of which is about $O(n^{1.16})$. Finally, we combine Cover Tree and KNNG to get TBSG, and adopt our edge selection strategy to make the graph close to a $(r,p)$-MSNET. The detail can be seen in Algorithm \ref{algo4}.

\subsection{Complexity Analysis}
The time cost for the construction of TBSG contains three parts: 1) establishing Cover Tree, 2) building an approximate KNNG, 3) building TBSG based on Cover Tree and approximate KNNG. The complexity of part 1) is $O(c^6n\log nd)$, where \textit{c} is the bounded expansion constant, \textit{n} is the cardinality and \textit{d} is the dimension. The cost of building Cover Tree is small in experiment. For part 2), we adopt the algorithm in Efanna\cite{fu2016efanna}, the complexity of which is $O(n^{1.16})$. The time cost of part 3) is mainly used for neighbor(edge) selection. According to Algorithm \ref{algo3}, the complexity of determining whether a candidate node is a neighbor is $O(md)$, where $m$ is the maximum of neighbor size. The total candidate set size is $2Kn$, where $K$ is the neighbor size of KNNG. Therefore, the complexity of part 3) is $O(Kmdn)$. The overall complexity of TBSG is $O(c^6n\log nd+n^{1.16}+Kmdn)$.

It is hard to estimate the search complexity of TBSG accurately, but according to our empirical study, the complexity is close to $O(\log nd)$. 

\begin{figure*}[!t]
\centering
\subfigure{\includegraphics[width=1.5in]{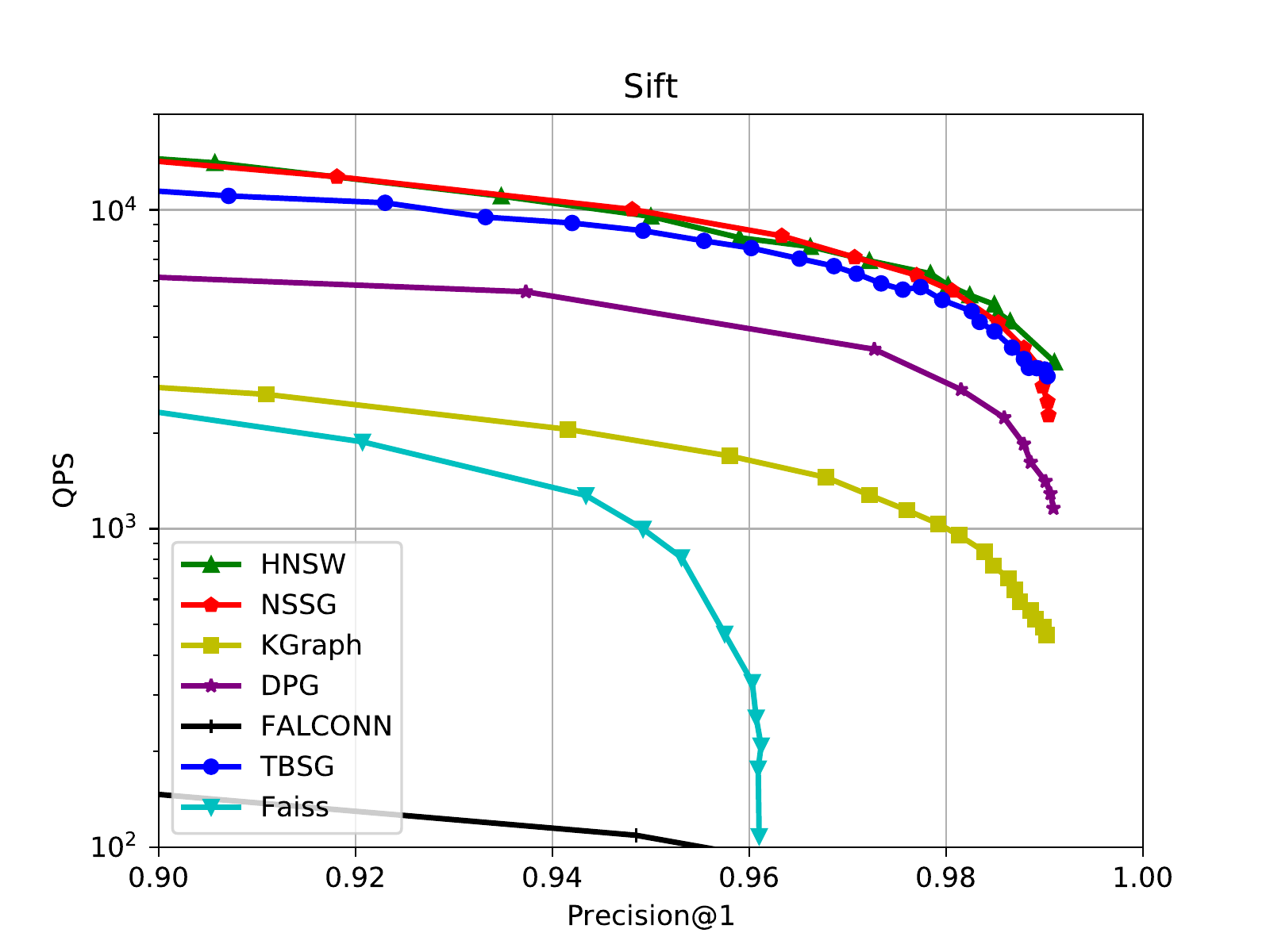}}
\subfigure{\includegraphics[width=1.5in]{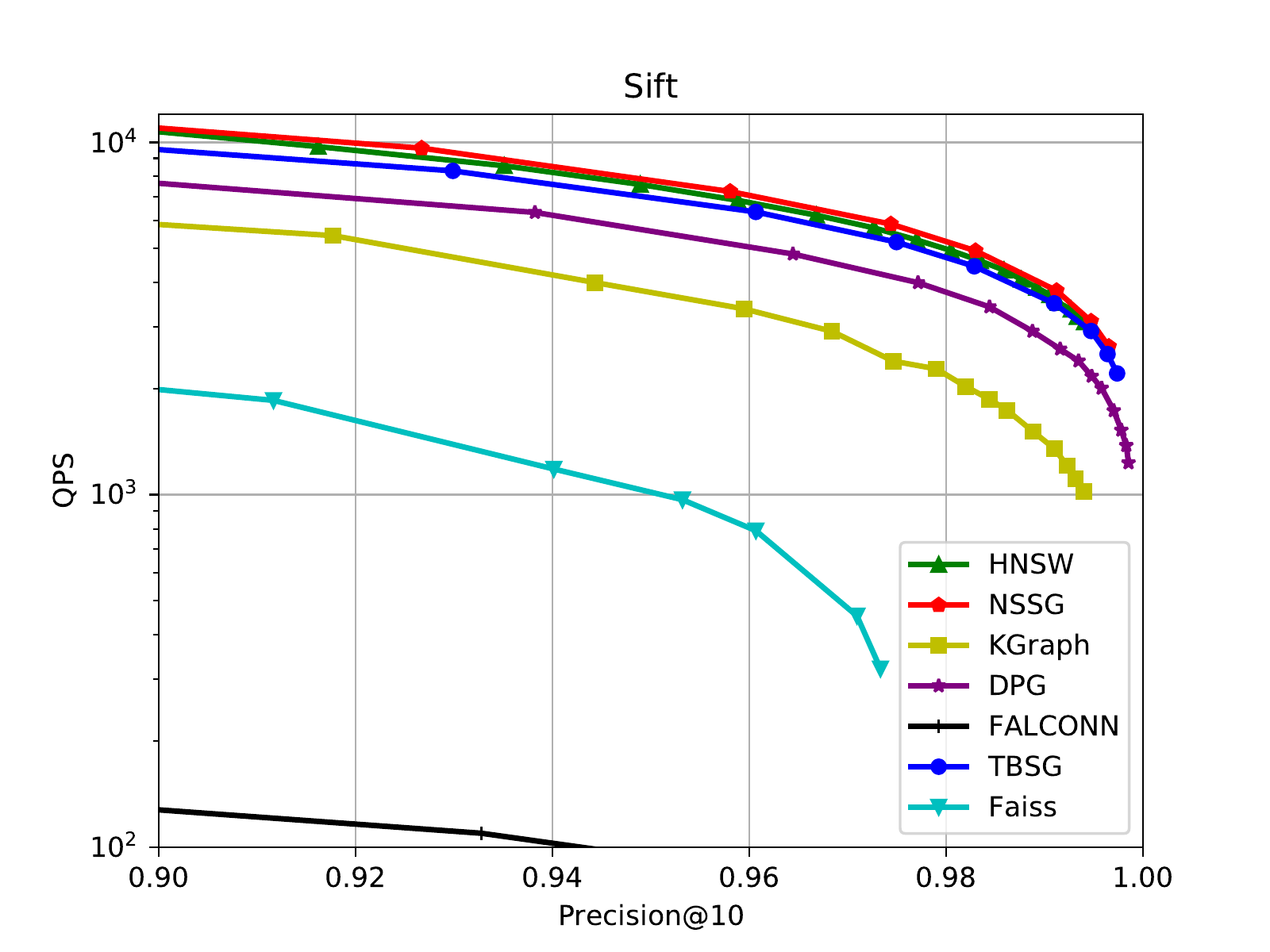}}
\subfigure{\includegraphics[width=1.5in]{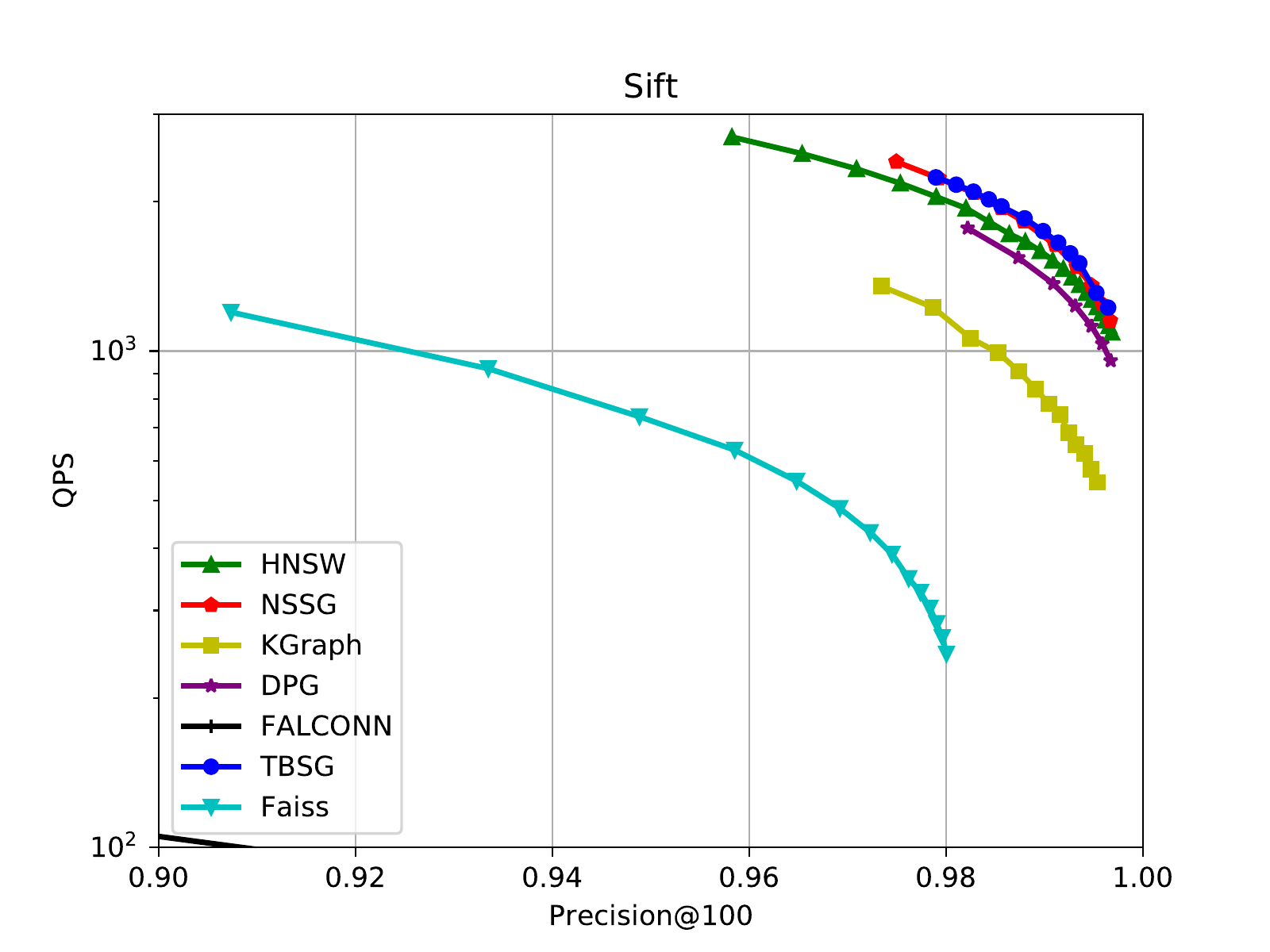}}
\subfigure{\includegraphics[width=1.5in]{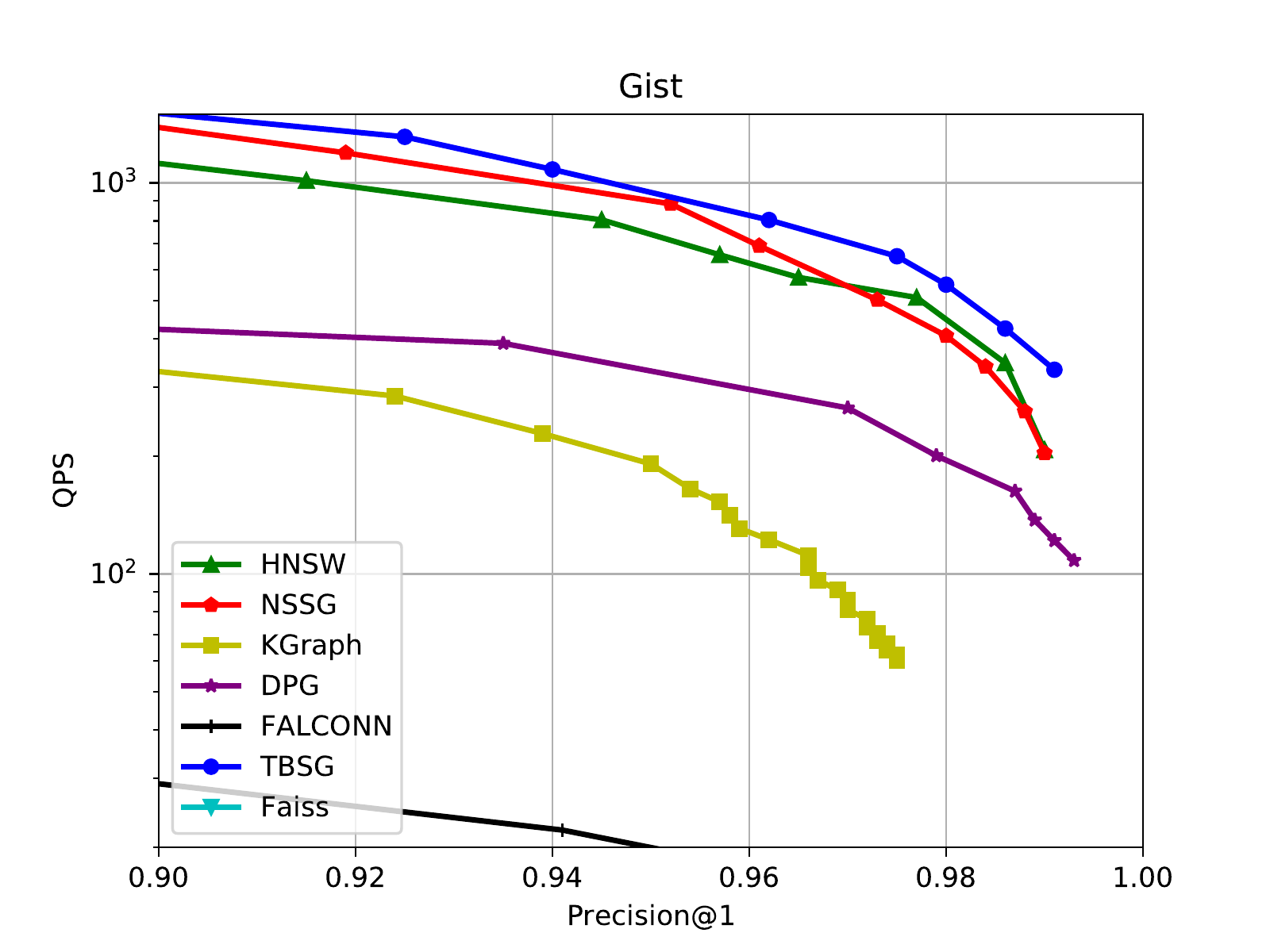}}
\subfigure{\includegraphics[width=1.5in]{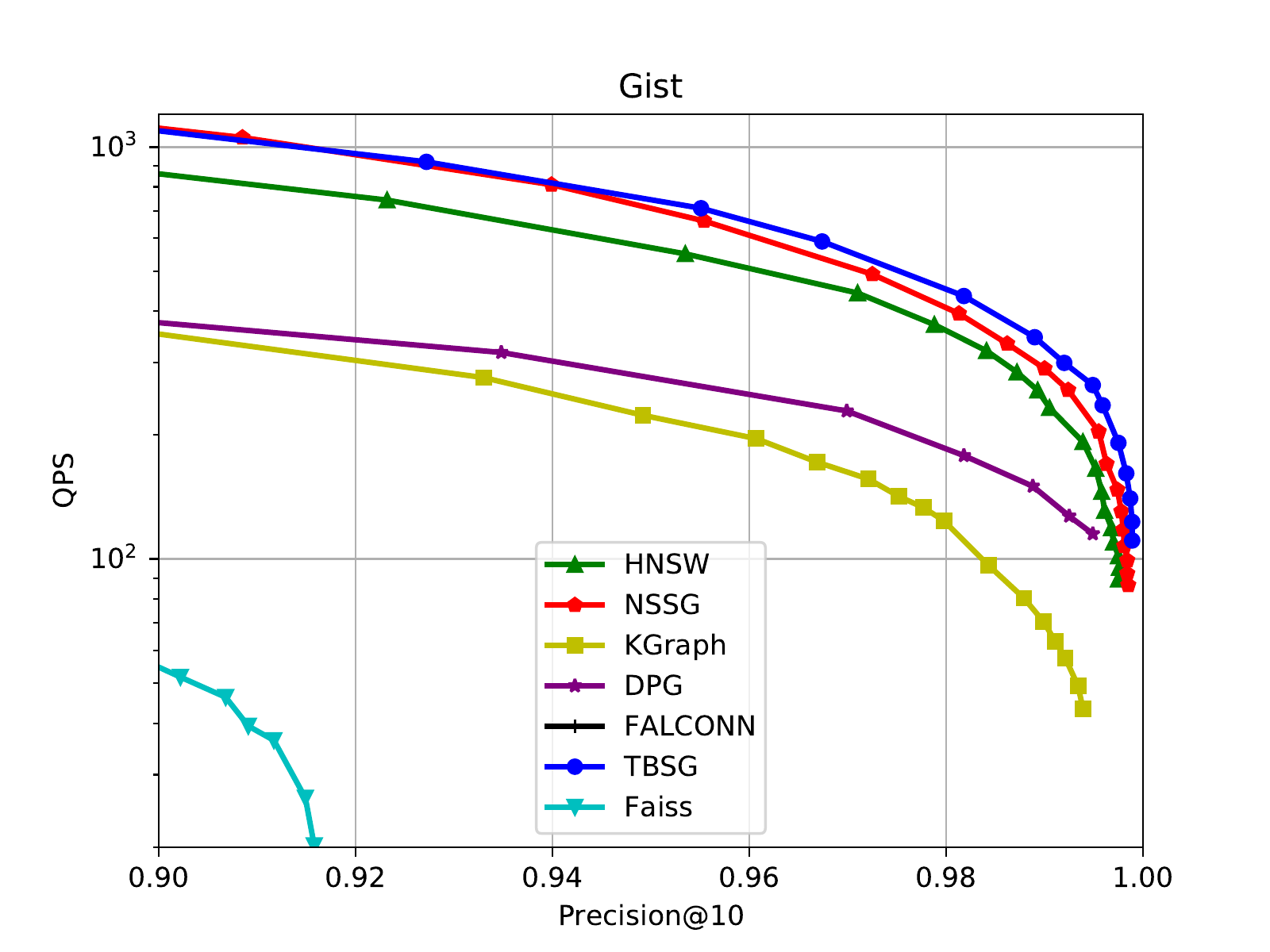}}
\subfigure{\includegraphics[width=1.5in]{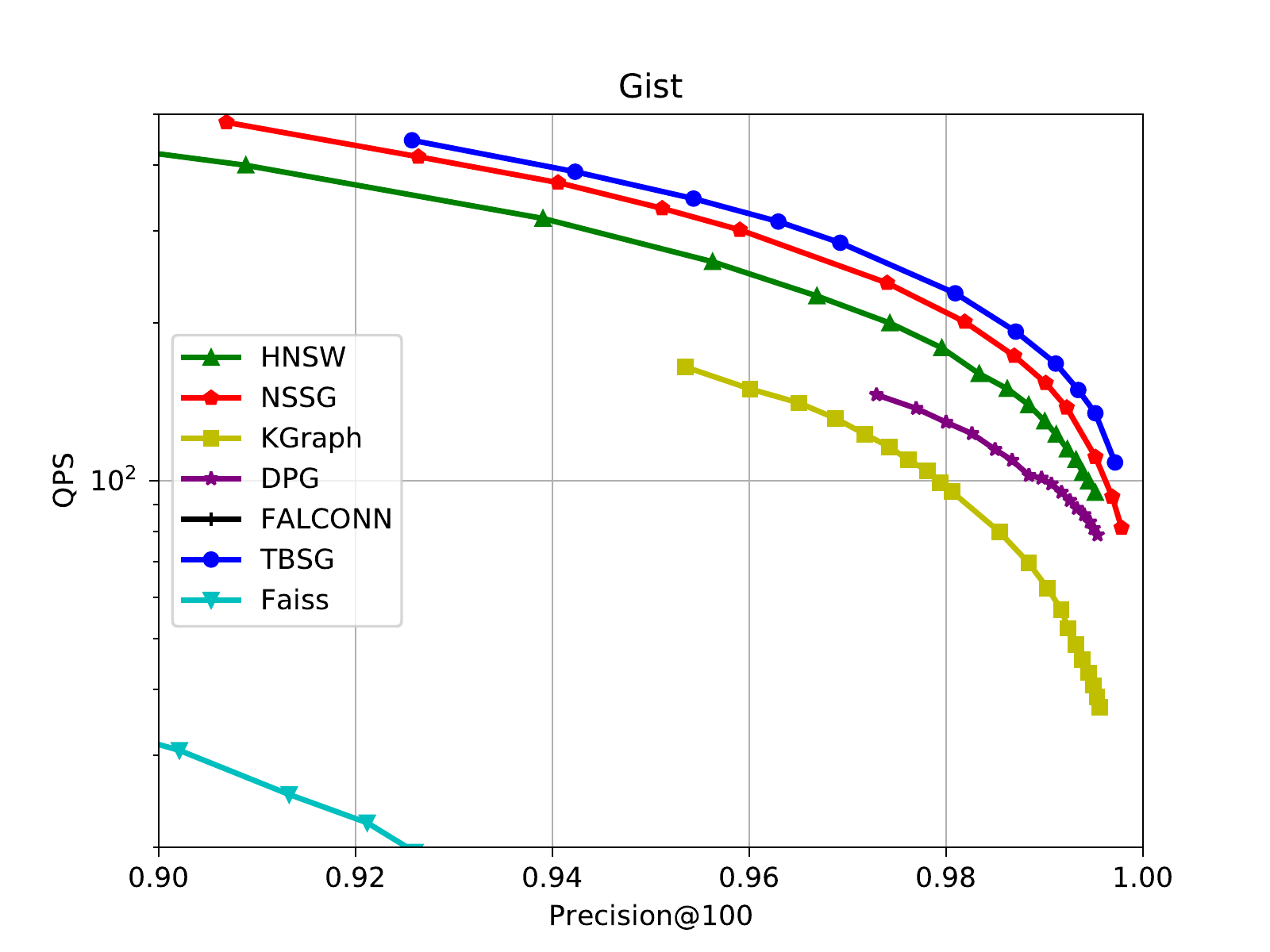}}
\subfigure{\includegraphics[width=1.5in]{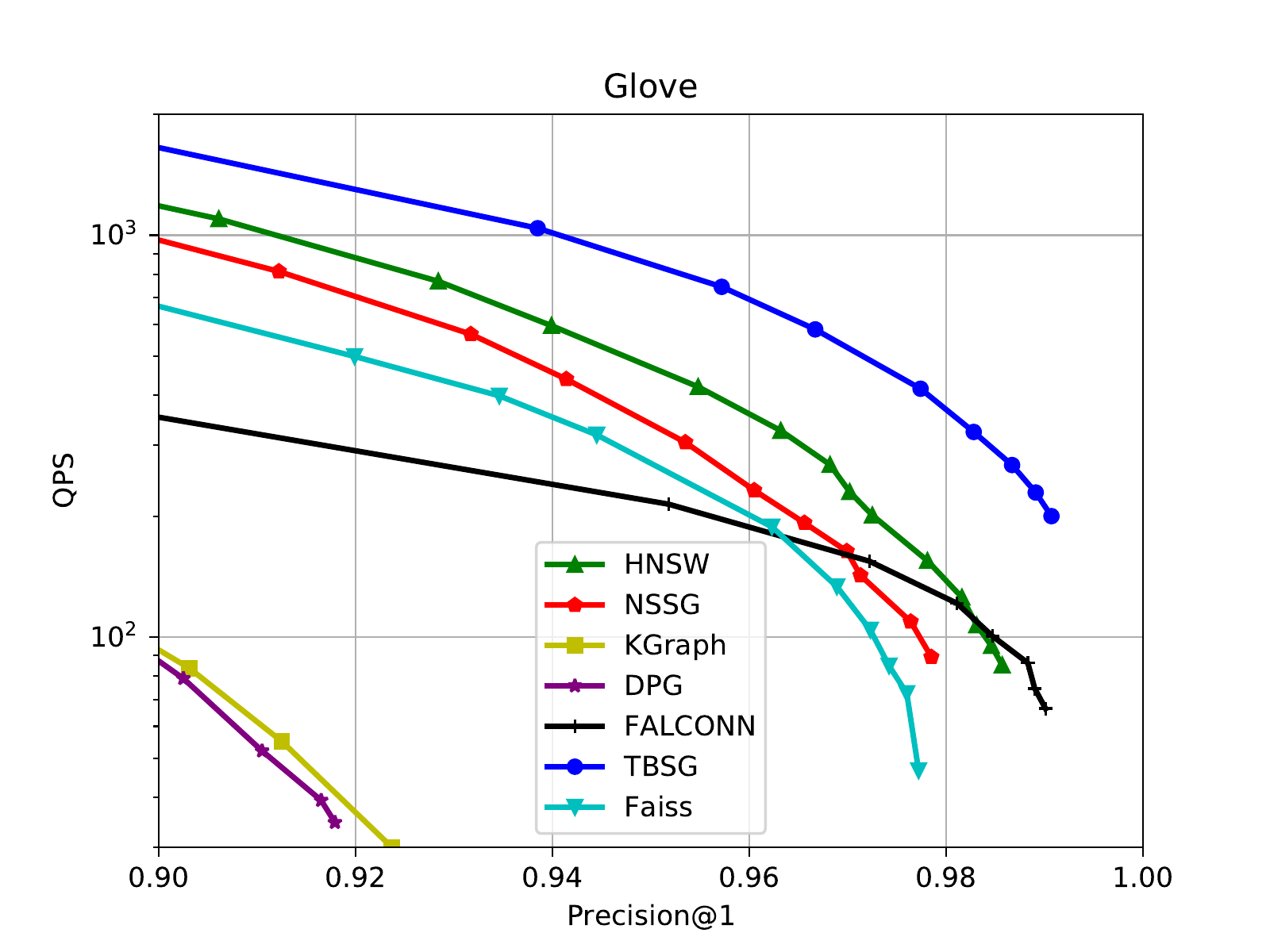}}
\subfigure{\includegraphics[width=1.5in]{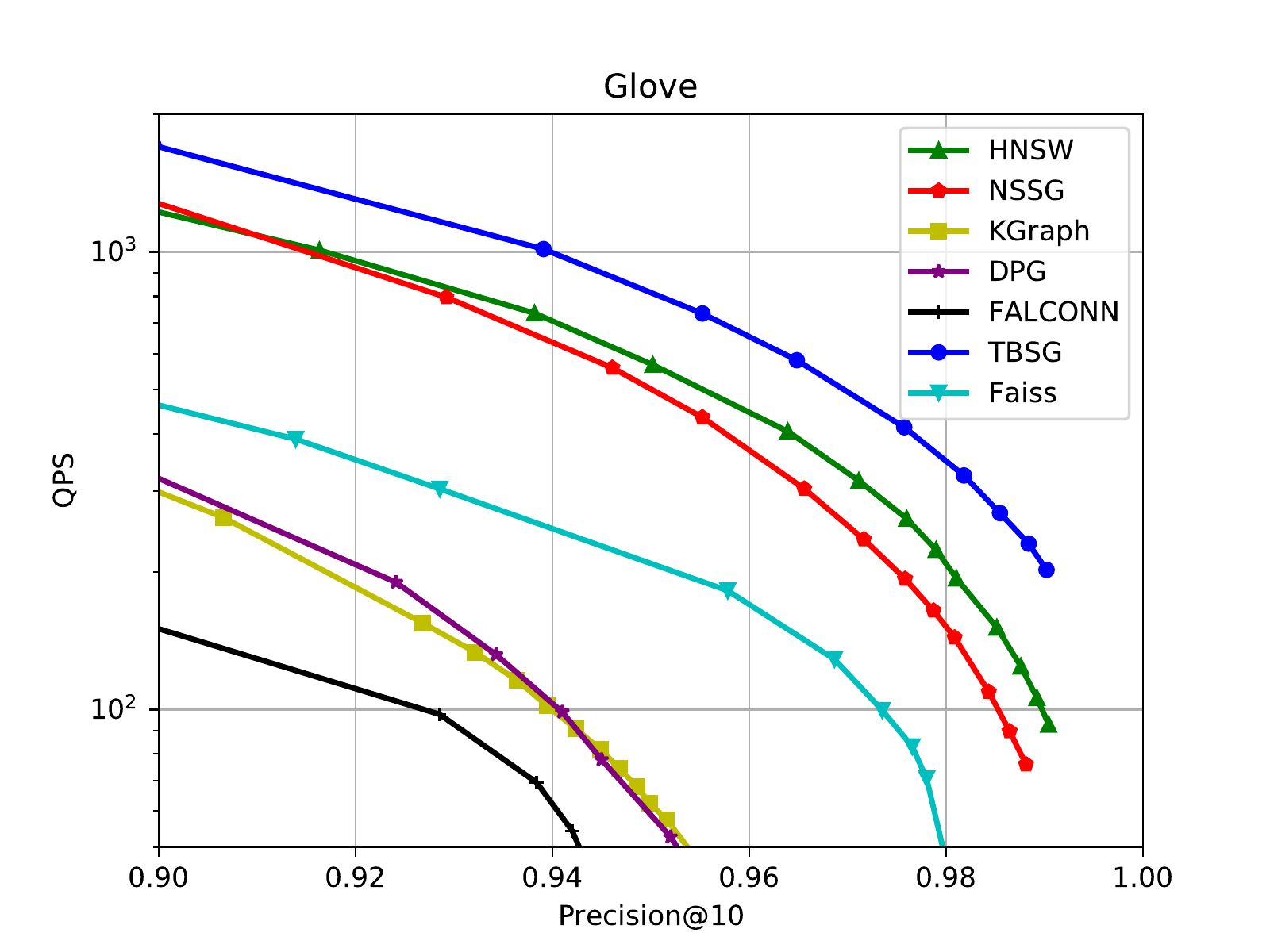}}
\subfigure{\includegraphics[width=1.5in]{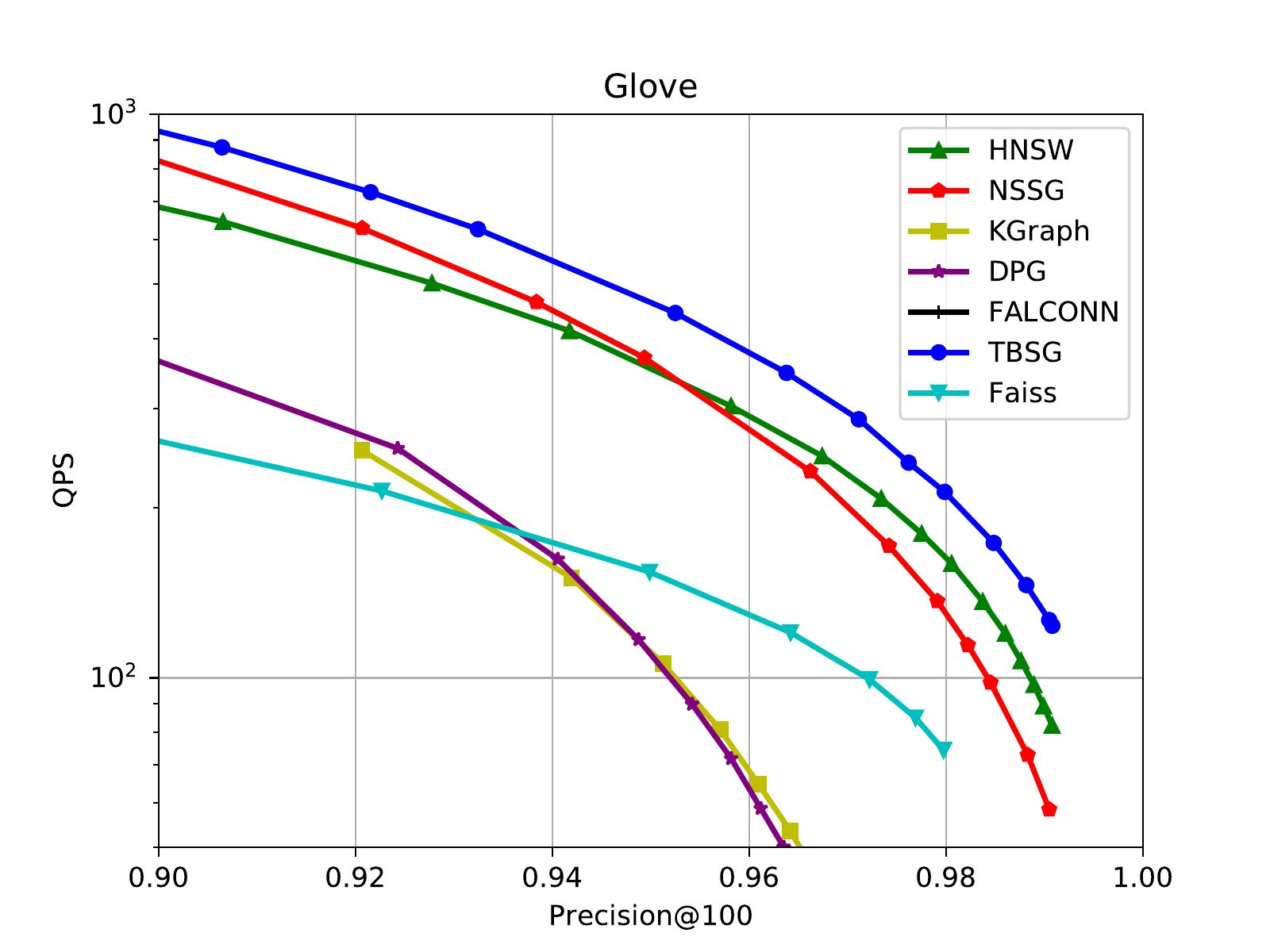}}
\subfigure{\includegraphics[width=1.5in]{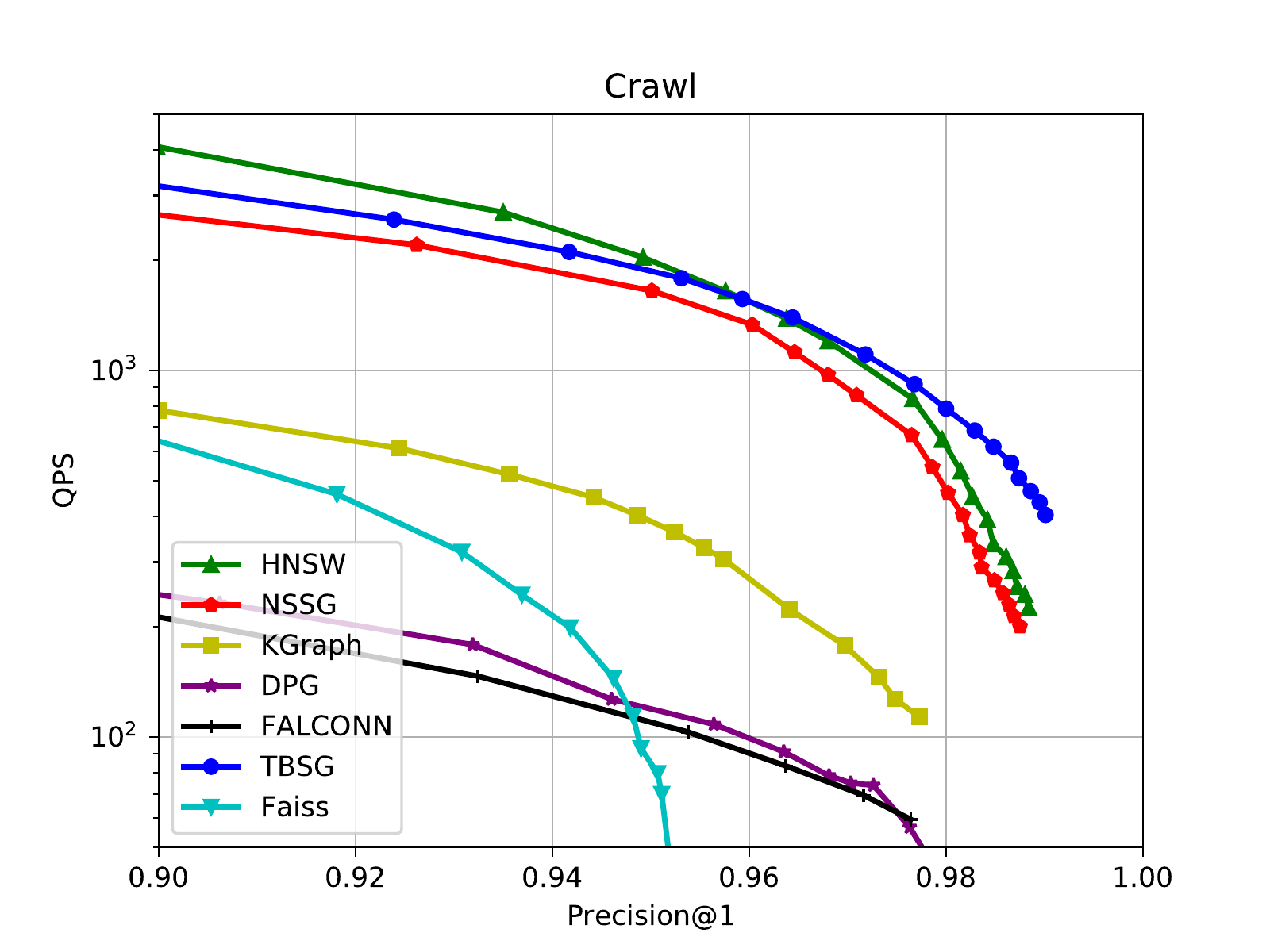}}
\subfigure{\includegraphics[width=1.5in]{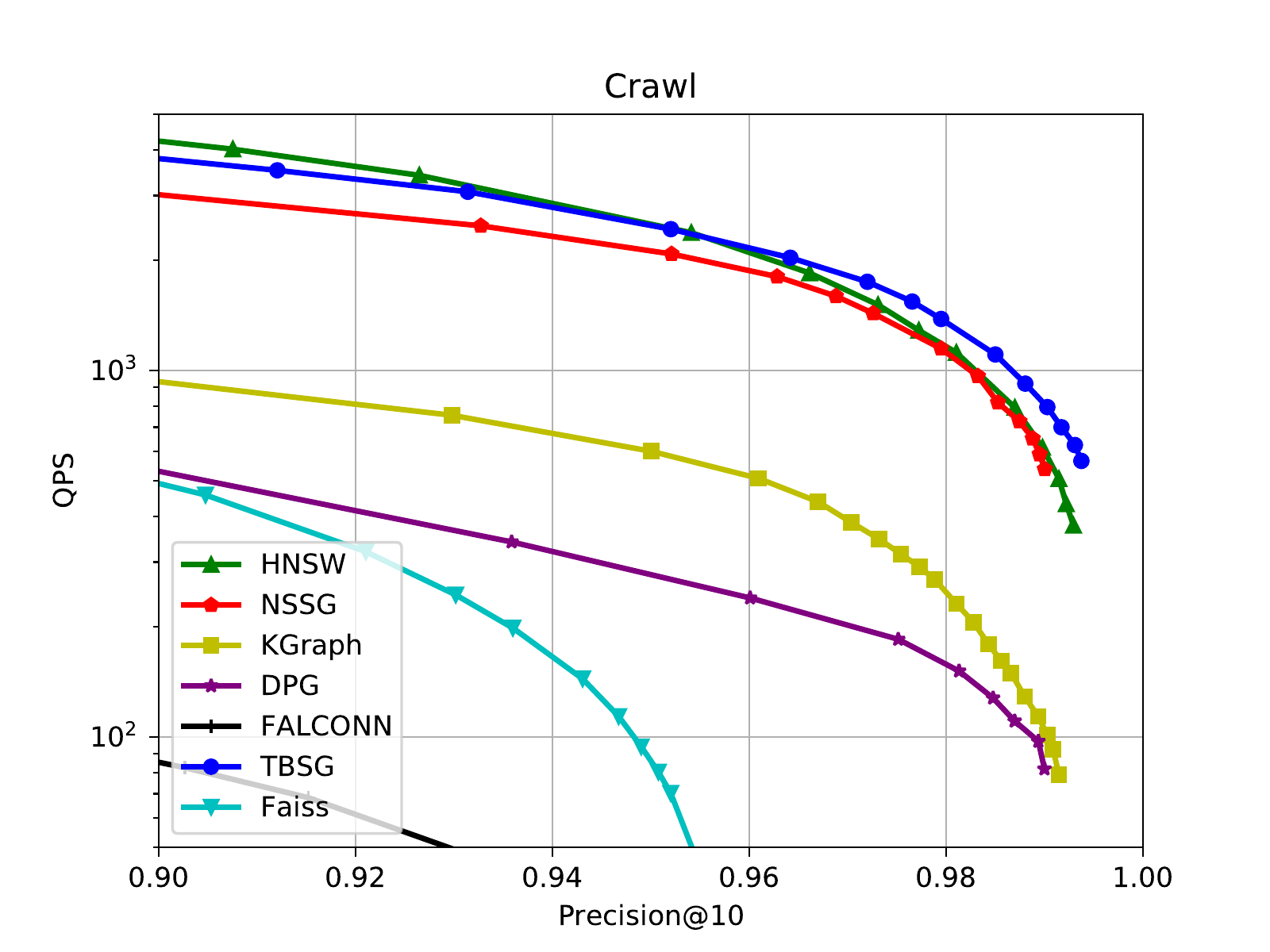}}
\subfigure{\includegraphics[width=1.5in]{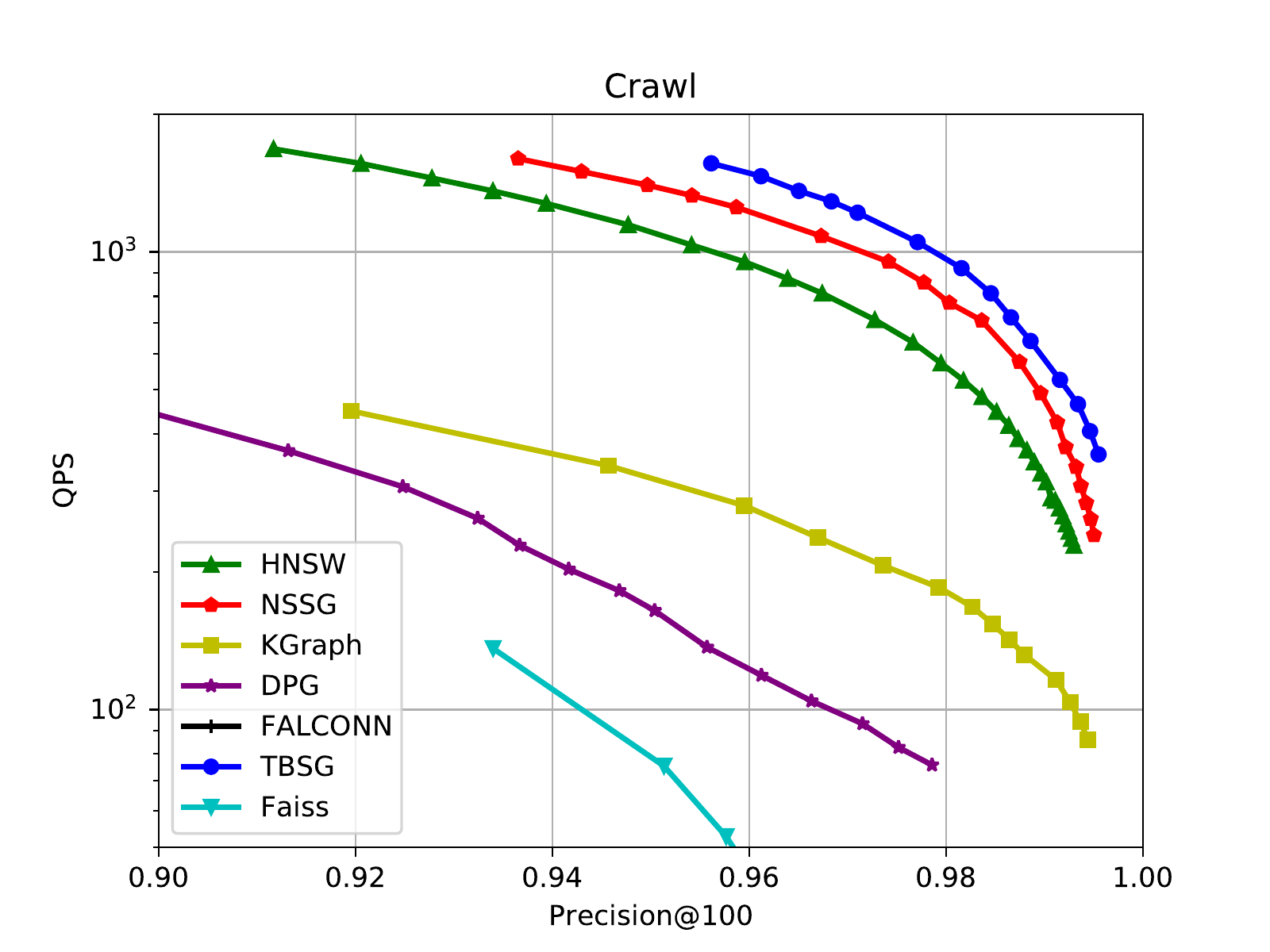}}
\caption{ The 1, 10, 100-NN search performance on 4 datasets.(top right is better)}
\label{fig:search_performance}
\end{figure*}

\section{Experiment}
In this part, we discuss the details of the experiment, including datasets, compared algorithms and the method for evaluation.
\subsection{Dataset}
We choose 4 million-scale datasets to evaluate: Sift\footnote{http://corpus-texmex.irisa.fr/}, Gist\footnote{http://corpus-texmex.irisa.fr/}, Glove\footnote{https://nlp.stanford.edu/projects/glove/}, Crawl\footnote{http://commoncrawl.org/}. Compared with their original dimension, we care more about their local intrinsic dimension (LID)\cite{costa2005estimating}. The detailed information of datasets is showed in Table \ref{table:dataset}.
\begin{table}
\centering
\caption{We list characteristics of the datasets including the number of data points(Nd) and query points(Nq), dimension(D), local intrinsic dimensionality (LID), where LID are used to describe the hardness of the datasets.}
\begin{tabular}{|c|c|c|c|c|} \hline
Dataset&Nd&Nq&D&LID\\ \hline
Sift & 1,000,000&10,000&128&9.3 \\ \hline
Gist& 1,000,000&1,000&960&18.9\\ \hline
Glove& 1,183,514&10,000&100&20.0\\ \hline
Crawl& 1,989,995&10,000&300&15.7\\
\hline\end{tabular}
\label{table:dataset}
\end{table}
\subsection{ TBSG v.s. SOTA methods}
\subsubsection{Evaluation}
We compare the search performance using queries-per-second (QPS) v.s. precision. The QPS is how much queries an algorithm can process per second at given precision. The precision is defined as \begin{displaymath}precision=\frac{\left |R\cap G\right |}{\left |G\right |}\end{displaymath}where $R$ is result returned by algorithm and $G$ is groundtruth.
\subsubsection{Compared Algorithms}
We select some current state-of-the-art algorithms for comparison, including tree based, product quantization based and graph based methods. We don't select the hash based methods because we're more concerned with efficiency at high precision, and they're too slow at high precision. We care more about the graph based algorithm, which are much more competitive. The algorithms to compare are as follows:
\begin{enumerate}[1.]
\item HNSW\footnote{https://github.com/nmslib/hnswlib}: Hierarchical NSW Graph is an improved version of the NSW Graph with superior search performance.
\item NSSG\footnote{https://github.com/ZJULearning/SSG}: Navigating SSG is an approximation of SSG with very low index cost, which is also current state-of-the-art algorithm. Also, NSSG is an improvement version of NSG.
\item KGraph\footnote{https://github.com/aaalgo/kgraph}: KGraph directly uses KNNG as the search index with a large value of K.
\item DPG\footnote{https://github.com/DBWangGroupUNSW/nns\_benchmark}: Diversified Proximity Graph selects neighbors by maximizing average pairwise angle from KNNG and adds reverse edges.
\item FALCONN\footnote{https://github.com/FALCONN-LIB/FALCONN} is a library with algorithms for ANNS, which are based on Locality-Sensitive Hashing (LSH).
\item Faiss\footnote{https://github.com/facebookresearch/faiss}: Faiss is a product quantization based algorithm released by Facebook. We used the IVF-PQ implementation for comparison.
\item TBSG\footnote{https://github.com/Fanxbin/TBSG}: Tree-based Search Graph is the algorithm proposed by this paper.
\end{enumerate}
All of the algorithms are implemented in C++, compiled by g++ with "O3" option. All the experiments are carried out on a machine with i7-10700K CPU @3.80GHZ*16 and 32 GB memory. We evaluate the search performance with a single thread and index performance with 16 threads.
\section{Result and Analysis}
\subsection{Search Performance}
We perform 1, 10, 100-NN queries with all the algorithms on four datasets, where 1, 10, 100 are the size of groundtruth per query. For graph-based methods, we adopt Algorithm \ref{alg1} for query and increase the result pool size to get higher precision.  Similarly, for Faiss we increase the parameter n\_probe and for FALCONN we increase the parameter num\_probes. We record the time cost when achieving the required precision and calculate the QPS. The search performance of all algorithms on 4 datasets is shown in Figure \ref{fig:search_performance}. We can see that:
\begin{enumerate}[1.]
\item  Our algorithm yields better or comparable search performance than all recent state-of-the-art algorithms.  In the case of datasets with larger LID and queries with higher precision, the advantages are more obvious; 
\item The performance of graph-based algorithms is much better than other algorithms in queries with high precision. In addition, the graph-based algorithms can achieve a higher precision limit. 
\item The edge-selecting strategies of DPG, NSSG and TBSG are all based on angle. In DPG, the degree of each node must exceed a given value (mostly very large), which results in the over large out-degree. However, both NSSG and TBSG can adjust the out-degree with hyper-parameter. Compared with NSSG considering only one angle, we consider both angles and quantify their impact on the probability of the monotonic search. NSSG' s advantage in Glove gradually disappears while TBSG' s advantage continues to expand, indicating that our design is more reasonable.
\end{enumerate}
\subsection{Index Performance}
\begin{table}
\centering
\caption{We list some important information of graph-based indexes, including the average out-degree(AOD), maximum out-degree(MOD), index size(Size)(MB) and index time(Time)(sec). Index size refers to the size of the index stored on the disk.}
\begin{tabular}{|c|c|c|c|c|c|}
\hline
Dataset&Algorithm&Size&AOD&MOD&Time\\
\hline
Sift & HNSW & 220.4&30&50 & 295 \\
\cline{2-6}
               &   KGraph  & 408& 100& 100 & 88 \\
\cline{2-6}
               &   DPG  & 191.3 & 45& 261& 88+30 \\
\cline{2-6}
               &   NSSG  & 160.7 & 39& 50& 115+75 \\
\cline{2-6}
               &   TBSG  & 174.7 & 40& 50& 115+80 \\
\hline

Gist & HNSW & 296&25&70 & 3021 \\
\cline{2-6}
               &   KGraph  & 1608& 400& 400 & 1202 \\
\cline{2-6}
               &   DPG  & 308 & 75& 7721& 1202+74\\
\cline{2-6}
               &   NSSG  & 143.5 & 34& 70& 1202+341 \\
\cline{2-6}
               &   TBSG  & 195 &44 & 70&1202+313  \\
\hline

Glove& HNSW & 254&18&50 & 1670 \\
\cline{2-6}
               &   KGraph  & 1903& 400& 400 & 905 \\
\cline{2-6}
               &   DPG  & 956 & 200& 200& 905+1370\\
\cline{2-6}
               &   NSSG  & 107 & 21& 50& 905+156 \\
\cline{2-6}
               &   TBSG  & 204 & 42& 80& 1198+119\\
\hline

Crawl & HNSW & 352&13&40 & 1952 \\
\cline{2-6}
               &   KGraph  & 3199& 400& 400 & 1190 \\
\cline{2-6}
               &   DPG  & 812 & 100& 100& 1190+651\\
\cline{2-6}
               &   NSSG  & 187 & 22& 40& 1190+380 \\
\cline{2-6}
               &   TBSG  & 221 & 26& 50& 1190+188 \\
\hline

\end{tabular}
\label{table:index_performance}
\end{table}

We list some important information of the graph-based indexes in Table \ref{table:index_performance}, including the average degree, maximum degree, index size and index time. In Glove and Crawl, the maximum degree of DPG is very large after adding reverse edges, which results in very low efficiency. So we undid the addition of reverse edges on these two datasets. According to Table \ref{table:index_performance}, we can see that:
\begin{enumerate}[1.]
\item The average out-degree of TBSG, HNSW and NSSG is much smaller than that of DPG and KGraph. This is mainly because in the edge-selecting strategy the out-degree of nodes is not fixed, and the maximum value is set to prevent over large out-degree. Both the search efficiency and precision are greatly affected by the average out-degree. When the average out-degree is too large, the precision will be higher but the search efficiency may decline. The average out-degree of TBSG is slightly higher than that of HNSW and NSSG, however, with the same precision the efficiency of TBSG is higher than that of HNSW and NSSG, indicating that TBSG can achieve a better trade-off between precision and efficiency.
\item The index time of TBSG is very close to that of NSSG, which is the lowest except for KGraph. For large-scaled datasets, it is very necessary to reduce index cost to improve the availability.
\item Index size refers to the size of the index stored on the disk. The index size of TBSG is smallest except NSSG.
\end{enumerate}
\subsection{Parameters}
The parameters to build TBSG are as follows: the size of neighbors in KNNG \textit{K}, the maximum of the out-degree \textit{m}, the threshold of $min\_prob$ \textit{mp}, the distance between query point and nearest neighbor \textit{r}. 

Generally speaking, it is harder to perform ANNS query in the dataset with a larger LID. Therefore, it needs more candidate neighbors and larger out-degree. The value of \textit{K} increases with LID, ranging from hundreds to thousands. The larger the value of \textit{K}, the more time it takes to build the index, and the more accurate the search result will be. The \textit{m} is set to prevent the over large out-degree. The value of \textit{m} is always within one hundred.

\begin{figure*}[!t]
\centering
\subfigure{\includegraphics[width=2.2in]{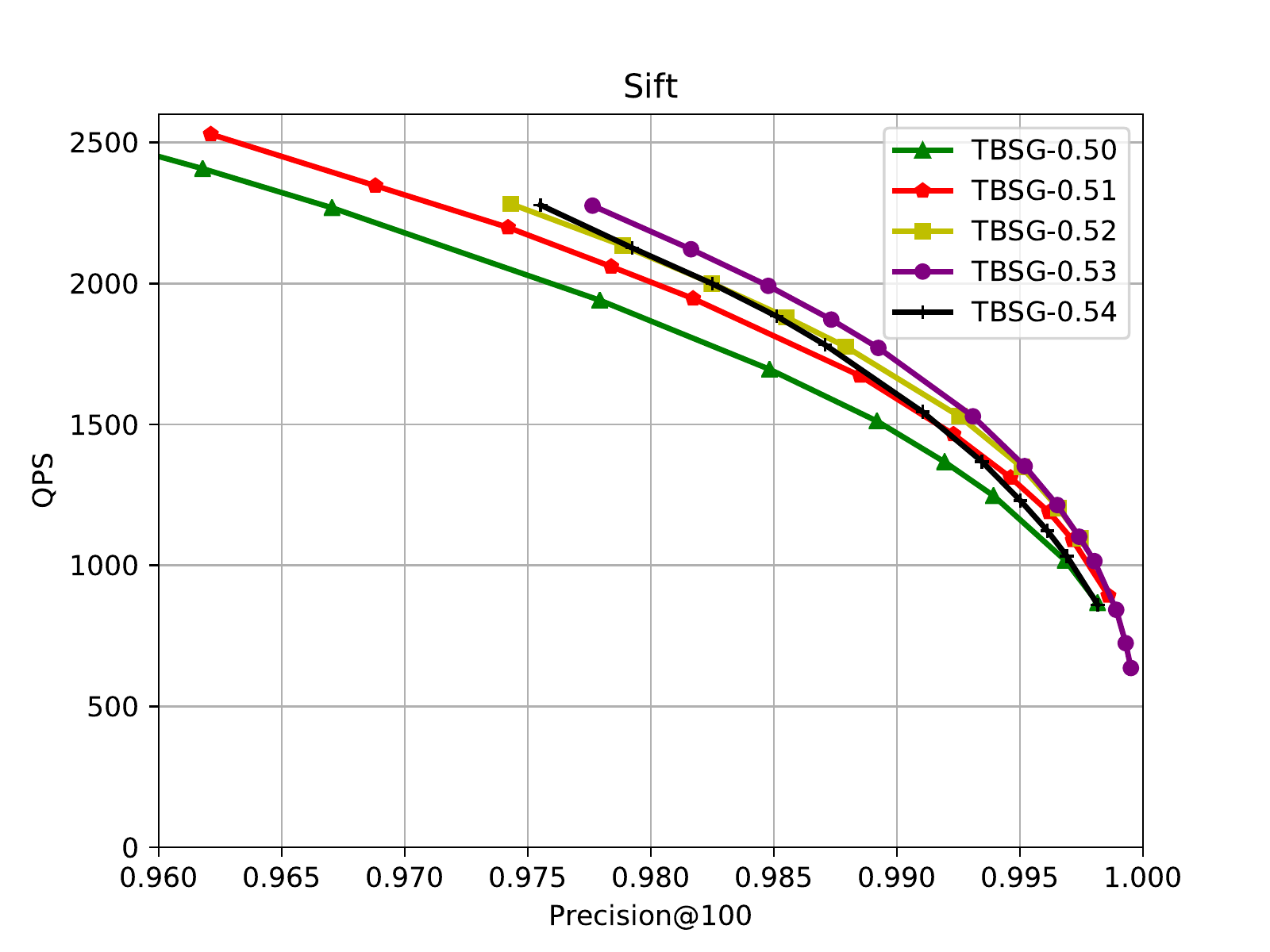}}
\subfigure{\includegraphics[width=2.2in]{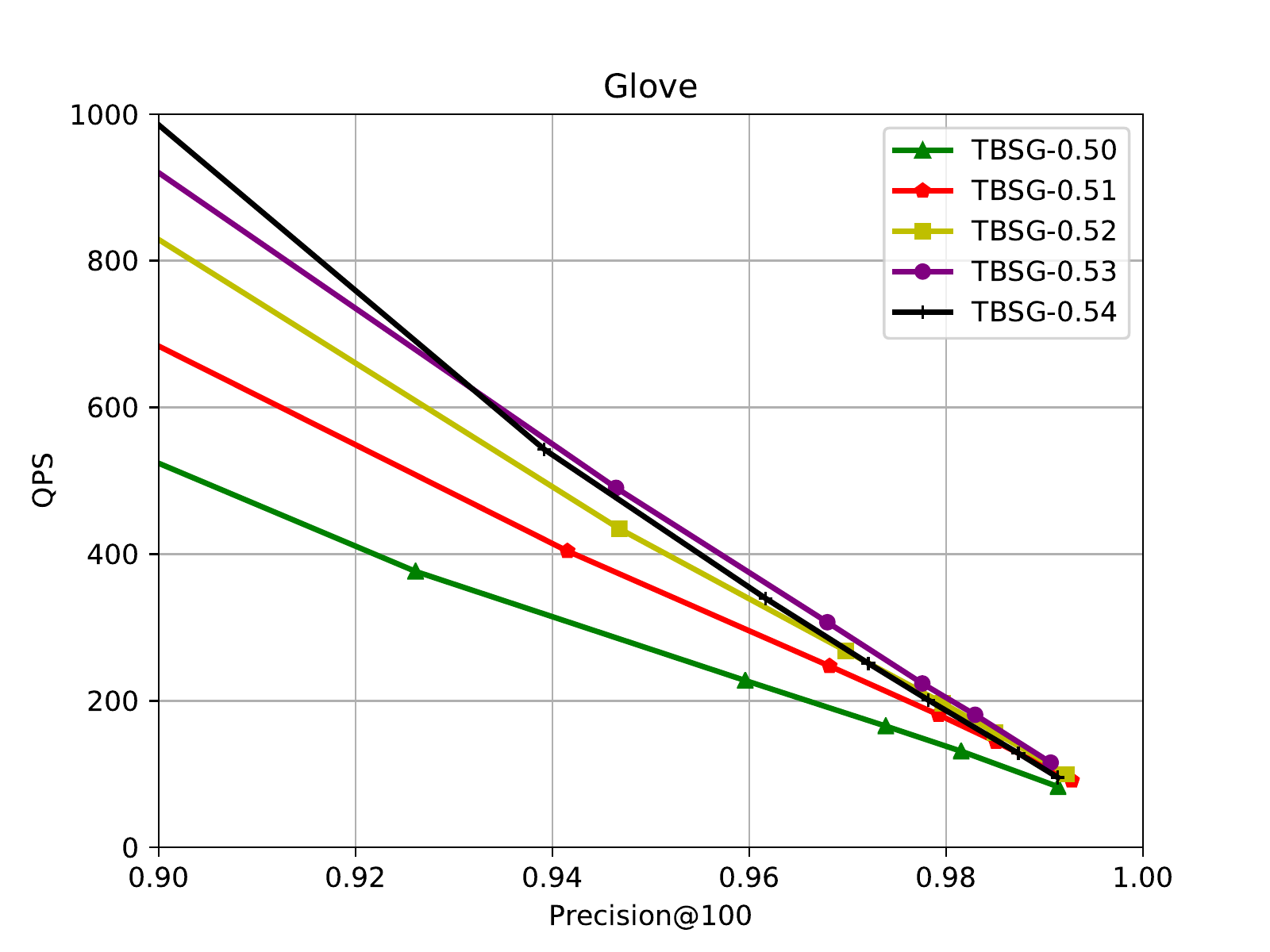}}
\caption{The result of using \textbf{$mp$} ranging from 0.5 to 0.54 on Sift and Glove.}
\label{fig:parameter}
\end{figure*}

There are two most important parameters in TBSG: \textit{r} and \textit{mp}. The value of \textit{r} can vary widely between different datasets, or between different nodes in the same dataset. Therefore, we consider two methods to determine the value of \textit{r}: static and dynamic. The static method is that for each node $v_i \in D$, we select a fixed value $r_i$. The value of $r_i$ is equal to the distance between $v_i$ and its nearest neighbor in $D$. The dynamic method is approximating \textit{r} as \textit{l}, the meaning of which can be seen in inequation (\ref{equa3}). In most experiments, we find that dynamic method performs better. Therefore, we adopt the second method in our experiment. Generally speaking, a larger value of \textit{mp} indicates a larger average out-degree, and the precision of search results is also higher. However, over large out-degree will also lead to the decline of efficiency. Therefore, with the increase of \textit{mp}, the efficiency will firstly increase and then decrease, which is confirmed by our experiment in Figure \ref{fig:parameter}. We only change \textit{mp} from 0.5 to 0.54 on Sift and Glove, and we can see that the maximum efficiency was achieved around 0.53.

The parameter setting of other algorithms is mainly based on \cite{fu2021nssg}, some of which are fine tuned for better performance.

\begin{figure*}[!t]
\centering
\subfigure{\includegraphics[width=2.2in]{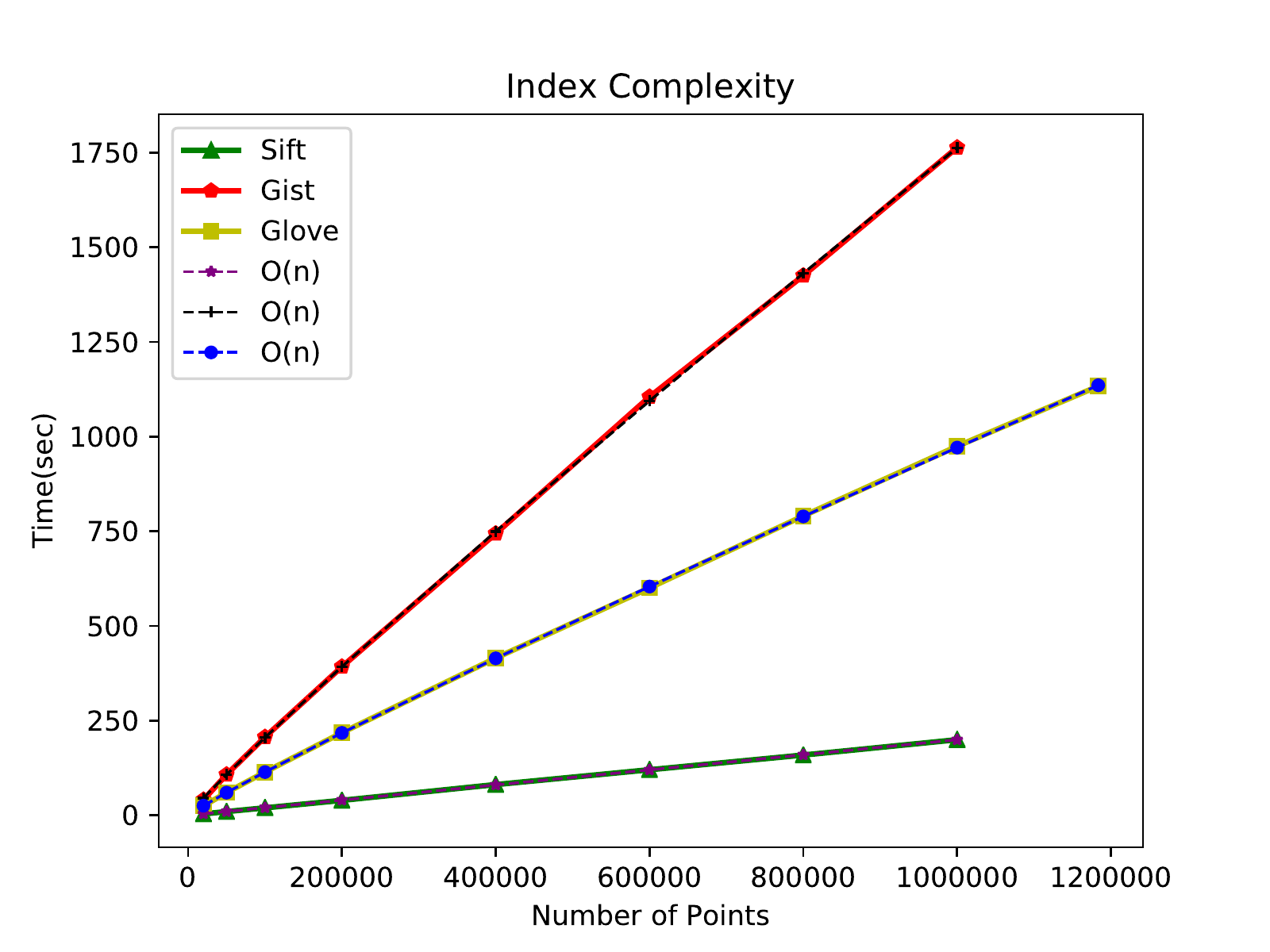}}
\subfigure{\includegraphics[width=2.2in]{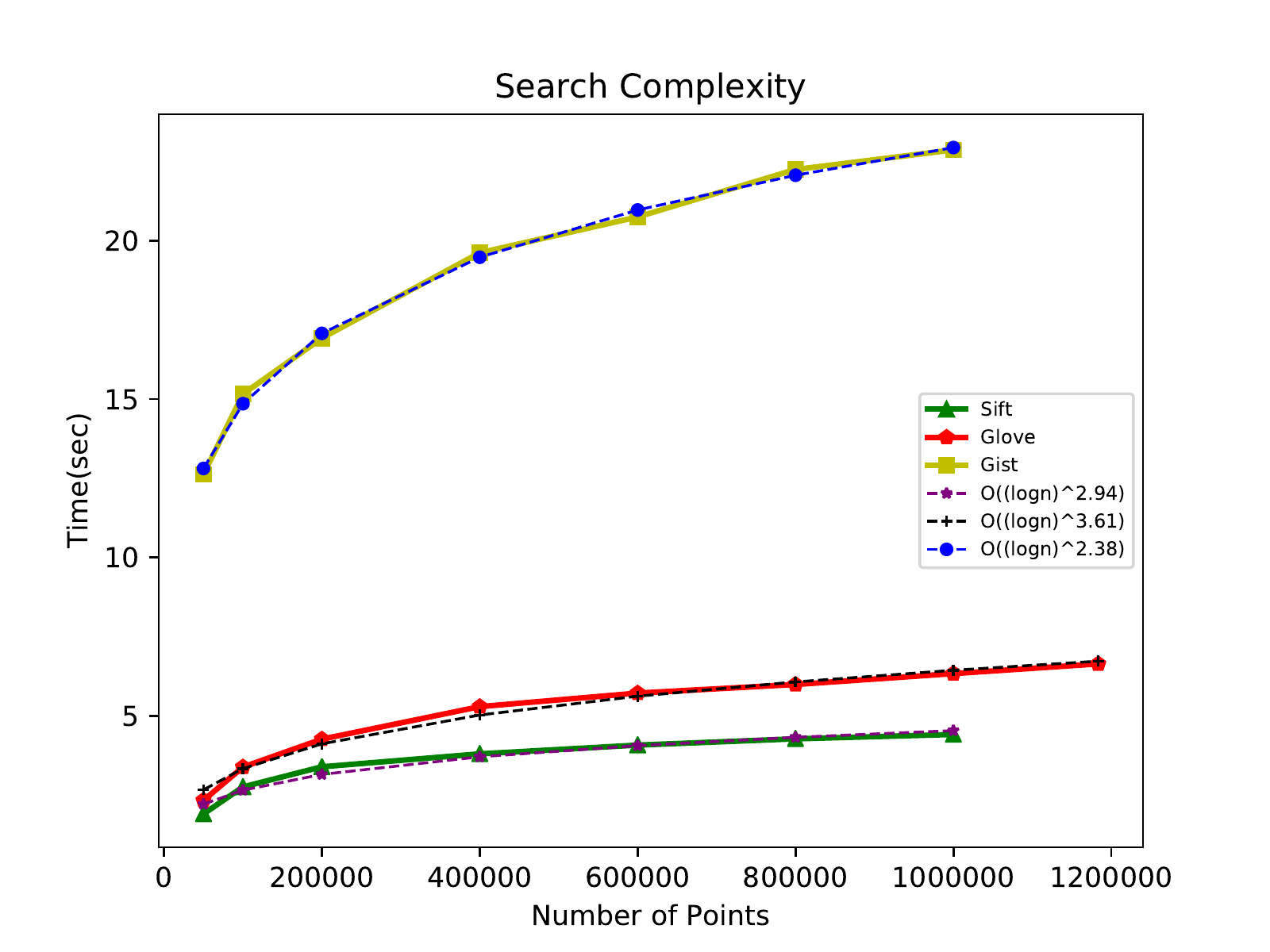}}
\caption{The result of index complexity and search complexity on Sift, Gist and Glove. }
\label{fig:complexity}
\end{figure*}
\subsection{Complexity}
In order to effectively evaluate the relationship between the index complexity and the size of dataset, we split the original datasets into subsets with different size and build indices on them with same parameters. Later, we use indexes above to conduct search using Algorithm \ref{alg1} with same query set and equal size of result pool. We carried out the above experiments on three datasets with very similar scales: Sift, Gist and Glove. We record the time cost for indexing and searching. The result is showed in Figure \ref{fig:complexity}.

The result shows that the index complexity of TBSG fits well with $O(n)$. The search complexity fits well with $O((\log n)^v)$ where \textit{v} is a small constant.
\section{Conclusions}
In this paper, we analyse and find that the current MSNET designs did not optimize the probability of the monotonic search, and the lower bound of the probability is only 50\%. To better match the query scenarios, we propose $(r,p)$-MSNET, which achieves monotonic search with the probability at least $p$ when the distance between query point and nearest neighbor is smaller than $r$. By adjusting the value of $p$, we can balance the length of search path and average out-degree of nodes to further improve the efficiency. Due to the high index complexity of a strict $(r,p)$-MSNET, we propose TBSG. Experiment conducted on several million-scaled datasets show that TBSG yields better or comparable search performance than all recent state-of-the-art algorithms.

\section{Appendix}
\subsection{Proof1}
We define the point set $\{Q: Q\in R_d  \cap \delta(B,Q) \leq r \cap \Pi_{\overrightarrow{AC}}{\overrightarrow{BQ}} >-r \cos{\frac{\varphi}{2}}  \}$ as $PS_1$, and the corresponding region as $Reg_1$. Define the point set $\{Q: Q\in R_d  \cap \delta(B,Q) \leq r \cap <\overrightarrow{AC},\overrightarrow{BQ}>  < \pi - \frac{\varphi}{2}\}$ as $PS_2$ where $<\overrightarrow{AC},\overrightarrow{BQ}>$ is the angle between $\overrightarrow{AC}$ and $\overrightarrow{BQ}$, and the corresponding region as $Reg_2$. Define the point set $\{Q:Q \in R_d \cap \delta(B,Q) < r\}$  as and the corresponding region as $Reg_3$.
We set up a cartesian coordinate system $x_1,x_2,\ldots,x_d$ in \textit{d} dimensional space with point \textit{B} as origin and specify that the positive direction of axis $x_1$ is same as $\overrightarrow{AC}$.
 The probability to achieve monotonic search is the ratio of the volume of $Reg_1$ to $Reg_3$, which can be represented as
 \begin{displaymath}Prob = \frac{Volume(Reg_1)}{Volume(Reg_3)}\end{displaymath}
 Because $PS_2 \in PS_1$, there is
 \begin{displaymath}Prob = \frac{Volume(Reg_1)}{Volume(Reg_3)}\end{displaymath} 
 \begin{displaymath}> \frac{Volume(Reg_2)}{Volume(Reg_3)}\end{displaymath}
 \begin{displaymath}=\frac{\int_{0}^{\pi-\frac{\varphi}{2}}\sin^{d-2}\theta_1 d\theta_1\int_{0}^{\pi}\sin^{d-3}\theta_2 d\theta_2 \ldots \int_{0}^{2\pi}d\theta_{d-1}\int_{0}^{r}R^{d-1}dR}{\int_{0}^{\pi}\sin^{d-2}\theta_1 d\theta_1\int_{0}^{\pi}\sin^{d-3}\theta_2 d\theta_2 \ldots \int_{0}^{2\pi}d\theta_{d-1}\int_{0}^{r}R^{d-1}dR}\end{displaymath}
 \begin{displaymath}=\frac{ \int_{0}^{\pi-\frac{\varphi}{2}}\sin^{d-2}\theta_1 d\theta_1 }{ \int_{0}^{\pi}\sin^{d-2}\theta_1 d\theta_1 }\end{displaymath}
 with \begin{align*}
\begin{split}
\left \{
\begin{array}{ll}
    x_1 = R\cos \theta_1,                    &\\
    x_2 = R \sin \theta_1\cos \theta_2,    & \\
    \ldots  &\\
    x_{d-1} = R \sin \theta_1 \ldots \sin \theta_{d-2} \cos \theta_{d-1} ,                             & \\
    x_{d} = R \sin \theta_1 \ldots \sin \theta_{d-2} \sin \theta_{d-1} 
\end{array}
\right.
\end{split}
\end{align*}
\begin{displaymath}\int_{0}^{\pi-\frac{\varphi}{2}}\sin^{d-2}\theta_1 d\theta_1=-\frac{\cos\theta_1\sin^{d-3}\theta_1}{d-2}\big|_{0}^{\pi-\frac{\varphi}{2}}\end{displaymath}
\begin{displaymath}+\frac{d-3}{d-2}\int_{0}^{\pi-\frac{\varphi}{2}}\sin^{d-4}\theta_1d\theta_1\end{displaymath}
because $\pi-\frac{\varphi}{2} \in [\frac{\pi}{2},\pi]$, there is
\begin{displaymath}\int_{0}^{\pi-\frac{\varphi}{2}}\sin^{d-2}\theta_1 d\theta_1 \ge  \frac{d-3}{d-2}\int_{0}^{\pi-\frac{\varphi}{2}}\sin^{d-4}\theta_1d\theta_1\end{displaymath}
\begin{displaymath}\int_{0}^{\pi}\sin^{d-2}\theta_1 d\theta_1=\frac{d-3}{d-2}\int_{0}^{\pi}\sin^{d-4}\theta_1d\theta_1 \end{displaymath}
Therefore, there is
\begin{displaymath}\frac{ \int_{0}^{\pi-\frac{\varphi}{2}}\sin^{d-2}\theta_1 d\theta_1 }{ \int_{0}^{\pi}\sin^{d-2}\theta_1 d\theta_1 } \ge \frac{\int_{0}^{\pi-\frac{\varphi}{2}}\sin^{d-4}\theta_1 d\theta_1 }{\int_{0}^{\pi}\sin^{d-4}\theta_1 d\theta_1 }\end{displaymath}
\begin{align*}
\begin{split}
\ge \left\{
\begin{array}{ll}
        1-\frac{\varphi}{2\pi}, d\quad is\quad even         &\\
        \frac{1+\cos \frac{\varphi}{2}}{2}, d\quad is\quad odd& 
\end{array}
\right.
\end{split}
\end{align*}
because $\frac{1+\cos \frac{\varphi}{2}}{2} \ge 1-\frac{\varphi}{2\pi}$, there is 
\begin{displaymath}Prob \ge 1-\frac{\varphi}{2\pi}\end{displaymath}
\subsection{parameter}
Here we list the value of parameters used for construction of indices. It is worth mentioning that the DPG is constructed based on KGraph and we cancel the addition of the reverse edges in Glove and Crawl.
\begin{enumerate}[1.]
\item Sift We use $k=100,mp=0.53, m=50$ for TBSG, $l=100, r=50,\alpha=60^\circ ,m=10$ for NSSG, $M=25,efconstruction=600$ for HNSW,$K=100,L=100,S=10,R=100,I=12$ for KGraph, $L_2=30$ for DPG, $num\_hash\_tables=50, num\_hash\_bits=18$ for FALCONN, IVF4096, PQ64+64 for Faiss.
\item Gist We use $k=200,mp=0.515, m=70$ for TBSG,$l=500, r=70,\alpha=60^\circ ,m=10$ for NSSG, $M=35,efconstruction=800$ for HNSW, $K=400,L=400,S=15,R=100,I=12$ for KGraph, $L_2=40$ for DPG, $num\_hash\_tables=50, num\_hash\_bits=18$ for FALCONN, IVF4096, PQ240+240 for Faiss.
\item Glove We use $k=300,mp=0.53, m=80$ for TBSG, $l=500, r=50,\alpha=60^\circ,m=10$ for NSSG, $M=40,efconstruction=2500$ for HNSW, $K=400,L=420,S=20,R=200,I=12$ for KGraph, $L_2=200$ for DPG,  $num_hash_tables=50, num\_hash\_bits=18$ for FALCONN,$num\_hash\_tables=50, num\_hash\_bits=18$ for FALCONN, IVF4096, PQ50+50 for Faiss.
\item We use $k=200,mp=0.53, m=50$ for TBSG, $l=500, r=40,\alpha=60^\circ,m=10$ for NSSG, $M=20,efconstruction=1000$ for HNSW, $K=400,L=420,S=15,R=100,I=12$ for KGraph, $L_2=100$ for DPG, $num\_hash\_tables=50, num\_hash\_bits=18$ for FALCONN, IVF4096, PQ100+100 for Faiss.
\end{enumerate}

\bibliography{main}

\begin{thebibliography}{10}
\expandafter\ifx\csname url\endcsname\relax
  \def\url#1{\texttt{#1}}\fi
\expandafter\ifx\csname urlprefix\endcsname\relax\def\urlprefix{URL }\fi
\expandafter\ifx\csname href\endcsname\relax
  \def\href#1#2{#2} \def\path#1{#1}\fi

\bibitem{arora2018hd}
A.~Arora, S.~Sinha, P.~Kumar, A.~Bhattacharya, Hd-index: pushing the
  scalability-accuracy boundary for approximate knn search in high-dimensional
  spaces, Proceedings of the VLDB Endowment 11~(8) (2018) 906--919.

\bibitem{beis1997shape}
J.~S. Beis, D.~G. Lowe, Shape indexing using approximate nearest-neighbour
  search in high-dimensional spaces, in: Proceedings of IEEE Computer Society
  Conference on Computer Vision and Pattern Recognition, IEEE Computer Society,
  1997, pp. 1000--1000.

\bibitem{zheng2016lazylsh}
Y.~Zheng, Q.~Guo, A.~K. Tung, S.~Wu, Lazylsh: Approximate nearest neighbor
  search for multiple distance functions with a single index, in: Proceedings
  of the 2016 International Conference on Management of Data, 2016, pp.
  2023--2037.

\bibitem{bentley1975multidimensional}
J.~L. Bentley, Multidimensional binary search trees used for associative
  searching, Communications of the ACM 18~(9) (1975) 509--517.

\bibitem{fukunaga1975branch}
K.~Fukunaga, P.~M. Narendra, A branch and bound algorithm for computing
  k-nearest neighbors, IEEE transactions on computers 100~(7) (1975) 750--753.

\bibitem{silpa2008optimised}
C.~Silpa-Anan, R.~Hartley, Optimised kd-trees for fast image descriptor
  matching, in: 2008 IEEE Conference on Computer Vision and Pattern
  Recognition, IEEE, 2008, pp. 1--8.

\bibitem{gionis1999similarity}
A.~Gionis, P.~Indyk, R.~Motwani, et~al., Similarity search in high dimensions
  via hashing, in: Vldb, Vol.~99, 1999, pp. 518--529.

\bibitem{huang2015query}
Q.~Huang, J.~Feng, Y.~Zhang, Q.~Fang, W.~Ng, Query-aware locality-sensitive
  hashing for approximate nearest neighbor search, Proceedings of the VLDB
  Endowment 9~(1) (2015) 1--12.

\bibitem{weiss2008spectral}
Y.~Weiss, A.~Torralba, R.~Fergus, et~al., Spectral hashing., in: Nips, Vol.~1,
  Citeseer, 2008, p.~4.

\bibitem{ge2013optimized}
T.~Ge, K.~He, Q.~Ke, J.~Sun, Optimized product quantization, IEEE transactions
  on pattern analysis and machine intelligence 36~(4) (2013) 744--755.

\bibitem{jegou2010product}
H.~Jegou, M.~Douze, C.~Schmid, Product quantization for nearest neighbor
  search, IEEE transactions on pattern analysis and machine intelligence 33~(1)
  (2010) 117--128.

\bibitem{zhang2014composite}
T.~Zhang, C.~Du, J.~Wang, Composite quantization for approximate nearest
  neighbor search, in: International Conference on Machine Learning, PMLR,
  2014, pp. 838--846.

\bibitem{harwood2016fanng}
B.~Harwood, T.~Drummond, Fanng: Fast approximate nearest neighbour graphs, in:
  Proceedings of the IEEE Conference on Computer Vision and Pattern
  Recognition, 2016, pp. 5713--5722.

\bibitem{fu2019NSG}
C.~Fu, C.~Xiang, C.~Wang, D.~Cai, Fast approximate nearest neighbor search with
  the navigating spreading-out graph, Proceedings of the VLDB Endowment 12~(5)
  (2019) 461--474.

\bibitem{hajebi2011fast}
K.~Hajebi, Y.~Abbasi-Yadkori, H.~Shahbazi, H.~Zhang, Fast approximate
  nearest-neighbor search with k-nearest neighbor graph, in: Twenty-Second
  International Joint Conference on Artificial Intelligence, 2011.

\bibitem{DPG}
W.~Li, Y.~Zhang, Y.~Sun, W.~Wang, M.~Li, W.~Zhang, X.~Lin, Approximate nearest
  neighbor search on high dimensional data-experiments, analyses, and
  improvement, IEEE Transactions on Knowledge and Data Engineering.

\bibitem{malkov2014NSW}
Y.~Malkov, A.~Ponomarenko, A.~Logvinov, V.~Krylov, Approximate nearest neighbor
  algorithm based on navigable small world graphs, Information Systems 45
  (2014) 61--68.

\bibitem{malkov2018HNSW}
Y.~A. Malkov, D.~A. Yashunin, Efficient and robust approximate nearest neighbor
  search using hierarchical navigable small world graphs, IEEE transactions on
  pattern analysis and machine intelligence.

\bibitem{cayton2008fast}
L.~Cayton, Fast nearest neighbor retrieval for bregman divergences, in:
  Proceedings of the 25th international conference on Machine learning, 2008,
  pp. 112--119.

\bibitem{fu2021nssg}
C.~Fu, C.~Wang, D.~Cai, High dimensional similarity search with satellite
  system graph: Efficiency, scalability, and unindexed query compatibility,
  IEEE Transactions on Pattern Analysis and Machine Intelligence.

\bibitem{kulis2009kernelized}
B.~Kulis, K.~Grauman, Kernelized locality-sensitive hashing for scalable image
  search, in: 2009 IEEE 12th international conference on computer vision, IEEE,
  2009, pp. 2130--2137.

\bibitem{wang2010sequential}
J.~Wang, S.~Kumar, S.-F. Chang, Sequential projection learning for hashing with
  compact codes.

\bibitem{liu2011hashing}
W.~Liu, J.~Wang, S.~Kumar, S.-F. Chang, Hashing with graphs, in: Icml, 2011.

\bibitem{jin2013density}
Z.~Jin, C.~Li, Y.~Lin, D.~Cai, Density sensitive hashing, IEEE transactions on
  cybernetics 44~(8) (2013) 1362--1371.

\bibitem{friedman1975algorithm}
J.~H. Friedman, J.~L. Bentley, R.~A. Finkel, An algorithm for finding best
  matches in logarithmic time, Department of Computer Science, Stanford
  University, 1975.

\bibitem{beckmann1990r}
N.~Beckmann, H.-P. Kriegel, R.~Schneider, B.~Seeger, The r*-tree: An efficient
  and robust access method for points and rectangles, in: Proceedings of the
  1990 ACM SIGMOD international conference on Management of data, 1990, pp.
  322--331.

\bibitem{2015Faster}
M.~Izbicki, C.~Shelton, Faster cover trees, European Journal of Operational
  Research 184~(3) (2015) 1179--1181.

\bibitem{beygelzimer2006cover}
A.~Beygelzimer, S.~Kakade, J.~Langford, Cover trees for nearest neighbor, in:
  Proceedings of the 23rd international conference on Machine learning, 2006,
  pp. 97--104.

\bibitem{babenko2014additive}
A.~Babenko, V.~Lempitsky, Additive quantization for extreme vector compression,
  in: Proceedings of the IEEE Conference on Computer Vision and Pattern
  Recognition, 2014, pp. 931--938.

\bibitem{boguna2009navigability}
M.~Boguna, D.~Krioukov, K.~C. Claffy, Navigability of complex networks, Nature
  Physics 5~(1) (2009) 74--80.

\bibitem{aurenhammer1991voronoi}
F.~Aurenhammer, Voronoi diagrams-a survey of a fundamental geometric data
  structure, ACM Computing Surveys (CSUR) 23~(3) (1991) 345--405.

\bibitem{dearholt1988monotonic}
D.~Dearholt, N.~Gonzales, G.~Kurup, Monotonic search networks for computer
  vision databases, in: Twenty-Second Asilomar Conference on Signals, Systems
  and Computers, Vol.~2, IEEE, 1988, pp. 548--553.

\bibitem{fu2016efanna}
C.~Fu, D.~Cai, Efanna: An extremely fast approximate nearest neighbor search
  algorithm based on knn graph, arXiv preprint arXiv:1609.07228.

\bibitem{costa2005estimating}
J.~A. Costa, A.~Girotra, A.~Hero, Estimating local intrinsic dimension with
  k-nearest neighbor graphs, in: IEEE/SP 13th Workshop on Statistical Signal
  Processing, 2005, IEEE, 2005, pp. 417--422.

\end{thebibliography}
\end{document}